\documentclass[aps,prx,twocolumn,english,superscriptaddress,floatfix,longbibliography]{revtex4-2}

\usepackage[english]{babel}
\usepackage{amsmath}
\usepackage{amssymb}
\usepackage{units}
\usepackage{upgreek}
\usepackage{braket}
\usepackage[colorlinks=true,urlcolor=blue,linkcolor=black,citecolor=black,bookmarksopen]{hyperref}
\usepackage{orcidlink}

\newcommand*{\finesse}{\mathcal{F}}

\begin{document}
\title{Spin-Photon Correlations from a Purcell-enhanced Diamond Nitrogen-Vacancy Center Coupled to an Open Microcavity}

\author{Julius~Fischer\orcidlink{0000-0003-2219-091X}}
\thanks{These authors contributed equally to this work.}
\author{Yanik~Herrmann\orcidlink{0000-0003-0545-7002}}
\thanks{These authors contributed equally to this work.}
\author{Cornelis~F.~J.~Wolfs\orcidlink{0009-0007-1776-3441}}
\author{Stijn~Scheijen\orcidlink{0009-0009-4092-4674}}
\author{Maximilian Ruf\orcidlink{0000-0001-9116-6214}}
\altaffiliation[Present address: ]{SandboxAQ, Palo Alto, California, USA}
\author{Ronald Hanson\orcidlink{0000-0001-8938-2137}}
\email{R.Hanson@tudelft.nl}
\affiliation{QuTech and Kavli Institute of Nanoscience, Delft University of Technology, P.O. Box 5046, 2628 CJ Delft, The Netherlands}
\date{\today}

\begin{abstract}
An efficient interface between a spin qubit and single photons is a key enabling system for quantum science and technology.
We report on a coherently controlled diamond nitrogen-vacancy center electron spin qubit that is optically interfaced with an open microcavity. Through Purcell enhancement and an asymmetric cavity design, we achieve efficient collection of resonant photons, while on-chip microwave lines allow for spin qubit control at a $\unit[10]{MHz}$ Rabi frequency. With the microcavity tuned to resonance with the nitrogen-vacancy center's optical transition, we use excited state lifetime measurements to determine a Purcell factor of $7.3 \pm 1.6$. Upon pulsed resonant excitation, we find a coherent photon detection probability of $\unit[0.5]{\%}$ per pulse. Although this result is limited by the finite excitation probability, it already presents an order of magnitude improvement over the solid immersion lens devices used in previous quantum network demonstrations.
Furthermore, we use resonant optical pulses to initialize and read out the electron spin. By combining the efficient interface with spin qubit control, we generate two-qubit and three-qubit spin-photon states and measure heralded Z-basis correlations between the photonic time-bin qubits and the spin qubit.
\end{abstract}

\maketitle

\section{Introduction}
Optically active solid-state qubits are promising physical systems for quantum networking \cite{kimble_quantum_2008,wehner_quantum_2018,ruf_quantum_2021,janitz_cavity_2020}. Among several candidates \cite{bernien_heralded_2013,knaut_entanglement_2024,ruskuc_multiplexed_2025,afzal_distributed_2024}, the nitrogen-vacancy (NV) center in diamond is one of the most-studied, with the realization of qubit teleportation \cite{hermans_qubit_2022} across a multi-node quantum network \cite{pompili_realization_2021} and metropolitan-scale heralded entanglement \cite{stolk_metropolitan-scale_2024}. Furthermore, efficient quantum frequency conversion to telecom wavelength \cite{tchebotareva_entanglement_2019,geus_low-noise_2024} and hybrid entanglement of photons and nuclear spins \cite{javadzade_efficient_2024,chang_hybrid_2025} have been demonstrated. The NV center offers excellent spin qubit properties \cite{abobeih_one-second_2018} and the nitrogen nuclear spin \cite{van_der_sar_decoherence-protected_2012} together with nearby carbon-13 spins \cite{childress_coherent_2006} can function as a multi-qubit register \cite{bradley_ten-qubit_2019,van_de_stolpe_mapping_2024} to store \cite{reiserer_robust_2016,bradley_robust_2022,bartling_entanglement_2022} and process \cite{taminiau_universal_2014,kalb_entanglement_2017,abobeih_fault-tolerant_2022} quantum states with high fidelities \cite{bartling_universal_2025}. The main challenges for the generation of spin-photon states with the NV center is the relatively low zero-phonon line (ZPL) emission of $\sim\unit[3]{\%}$ \cite{riedel_deterministic_2017} and the reduced optical coherence in the presence of charge noise hindering integration into nanophotonic devices \cite{faraon_coupling_2012,ishikawa_optical_2012,lekavicius_diamond_2019,jung_spin_2019}. State-of-the-art quantum network nodes \cite{hermans_qubit_2022} are thus realized with solid immersion lenses (SIL) \cite{hadden_strongly_2010}, in which the NV center retains lifetime-limited optical linewidth \cite{hermans_entangling_2023}. In practice, such systems have shown ZPL detector click probabilities around $5 \times 10^{-4}$ upon pulsed resonant excitation \cite{hensen_loophole-free_2015,humphreys_deterministic_2018,pompili_realization_2021,hermans_qubit_2022}.

In this work, we realize the integration of a NV center into a fiber-based Fabry-P\'{e}rot microcavity \cite{vahala_optical_2003,hunger_fiber_2010} in combination with coherent microwave spin control, enabling the generation of spin-photon states. This platform features bulk-like optical coherence of the NV center by incorporating a micrometer-thin diamond membrane into the cavity \cite{ruf_optically_2019,ruf_resonant_2021} and a Purcell-enhanced emission into the ZPL combined with an efficient photon extraction. We quantify the coupling of the NV center's readout transition to the cavity modes with the Purcell factor by measuring the reduction of the excited state lifetime. Further, we investigate the cavity-coupled readout transition with a photoluminescence excitation (PLE) measurement to determine the optical linewidth, a Hanbury Brown and Twiss (HBT) experiment to verify single-photon emission, and a saturation measurement, which reveals a ZPL detector click probability that is an order of magnitude higher compared to standard solid immersion lens systems. Moreover, we demonstrate coherent control of the NV center's electron spin and characterize its coherence properties in a Ramsey and a Hahn-Echo measurement. Finally, we use our platform to generate spin-photon states of the electron spin qubit with one and two time-bin encoded ZPL photon qubits. We herald on the photon detection in their time-bin states and observe the correlation with the corresponding spin qubit readout.

\begin{center}
\begin{figure*}[ht]
    \includegraphics[scale=1.6]{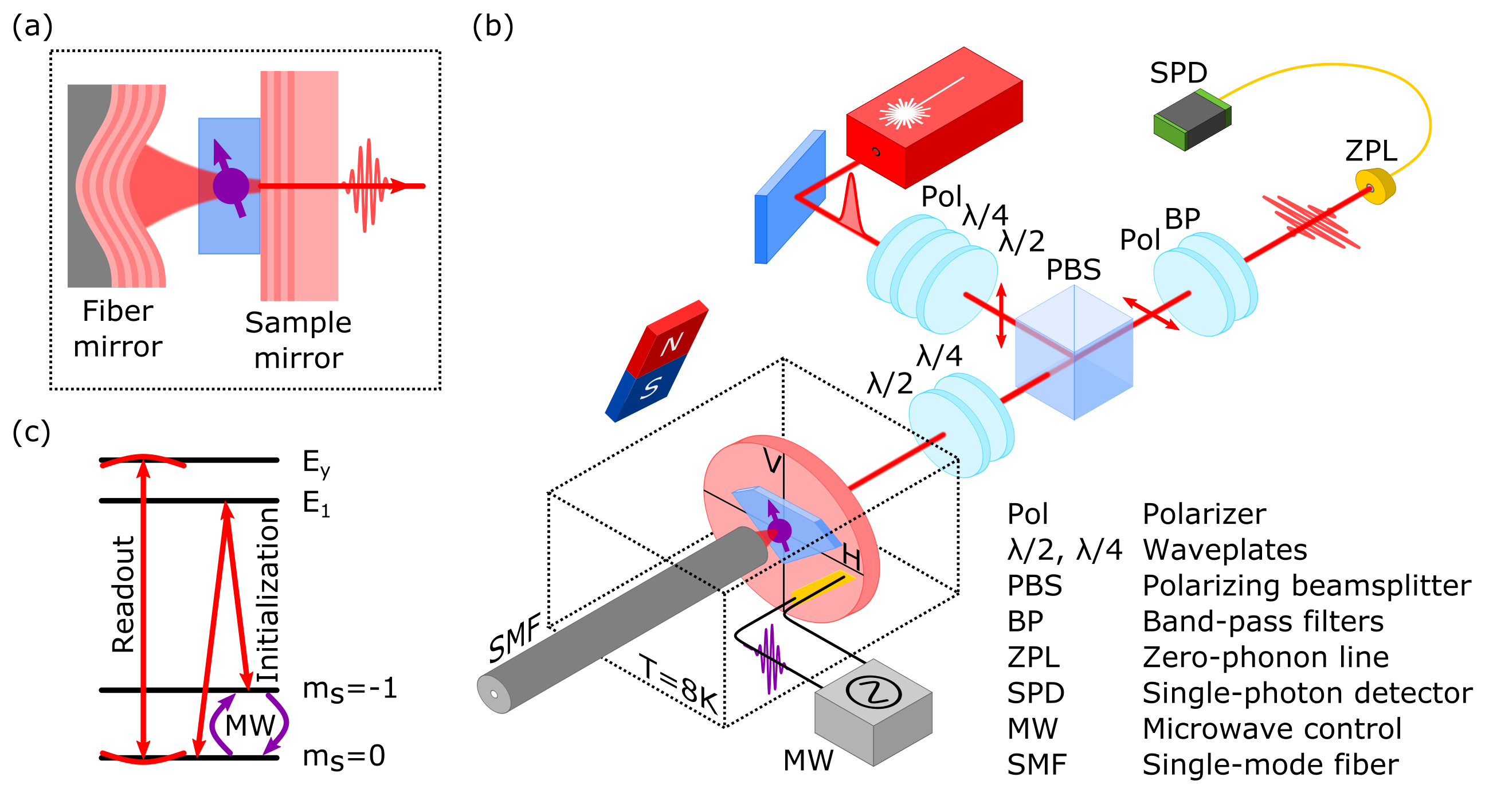}
    \caption{(a) Schematic of the fiber-based Fabry-P\'{e}rot microcavity. The NV center is hosted in a diamond membrane bonded to a planar sample mirror that faces a spherical mirror on the tip of an optical fiber, forming the cavity. The cavity mirrors are dielectric Bragg mirrors. (b) Schematic experimental setup to efficiently interface a NV center qubit with an optical microcavity. The cavity-coupled NV center spin qubit is controlled with microwave signals, which are delivered via a gold stripline that is embedded into the sample mirror. Outside the cryostat, a permanent magnet is mounted to apply a static magnetic field to the qubit. The optical readout of the cavity-coupled NV center is achieved using free-space cross-polarized resonant excitation and detection via the sample mirror side. Additional lasers for charge repump, spin initialization, and cavity locking are deployed in fiber via the fiber mirror side. (c) Energy diagram showing the relevant levels of the cavity-coupled NV center. The microwave qubit control, as well as the optical interface for readout and spin initialization, is depicted. The cavity is on resonance with the NV center $E_y$ transition for readout.}
    \label{fig:microcavity_schematic}
\end{figure*}
\end{center}

\section{Interfacing Diamond Nitrogen-Vacancy Center Spin Qubits with an Optical Microcavity}
In this section, we conceptually outline the optical interfacing of NV centers which are coupled to a fiber-based Fabry-P\'{e}rot microcavity as schematically depicted in Fig.~\ref{fig:microcavity_schematic}(a). At its heart, a sample mirror with a bonded diamond membrane faces a laser-ablated spherical mirror on the tip of an optical fiber forming the microcavity \cite{hunger_fiber_2010}. Using the experimental setup presented in Fig.~\ref{fig:microcavity_schematic}(b), the cavity-coupled NV centers are optically addressed via the microcavity, which is mounted inside a closed-cycle optical cryostat. Details about the low-vibration cryogenic microcavity setup can be found in Ref.~\cite{herrmann_low-temperature_2024}, and the fabrication of the diamond device, which is also used in earlier work \cite{ruf_resonant_2021}, is reported in Ref.~\cite{ruf_optically_2019}.

For the experiments in this study, the two orthogonal, linear polarization modes of the microcavity must be considered. Due to birefringence in the diamond membrane, the frequency degeneracy of these cavity modes is lifted, which is identified as a polarization mode splitting in the optical cavity response. In our lab frame, the low frequency (LF) and high frequency (HF) cavity mode is referred to as horizontal and vertical polarization cavity modes, respectively. The NV center's optical dipole of the readout transition around $\unit[637]{nm}$ couples to both polarization cavity modes, enabling resonant excitation and detection in a cross-polarized fashion \cite{yurgens_cavity-assisted_2024}. Nanosecond-short optical excitation pulses are coupled into the vertical polarization cavity mode through the sample mirror via the reflection of a polarizing beamsplitter. Upon excitation, the NV center ZPL emission into the horizontal polarization cavity mode is coupled out through the sample mirror and detected by a fiber-coupled single-photon detector in transmission of the polarizing beamsplitter. This cross-polarization filtering reaches high suppression values of the excitation laser and is actively optimized during the experiments via piezo rotation mounted waveplates. In addition, a cavity lock laser at $\unit[637]{nm}$ is launched over the optical fiber mirror into the horizontal polarization cavity mode and is detected in cavity transmission by the same single-photon detector. This signal is used in the experiments to run a side-of-fringe lock that feeds back on the piezo that is controlling the fiber mirror position to maintain a constant cavity length. More details about the experimental setup and the methods to maintain a high excitation laser suppression and to lock the cavity are described in Appendix~\ref{apx:setup}.

To prepare the NV center in its negatively charged state an off-resonant $\unit[515]{nm}$ charge repump laser is deployed over the fiber mirror next to a second, $\unit[637]{nm}$ laser to initialize the NV center in its $m_s=0$ spin ground state via the $E_1$ transition (see Fig.~\ref{fig:microcavity_schematic}(c)). A permanent magnet outside the cryostat creates a static magnetic field at the position of the NV center, which Zeeman-splits the NV center's $m_s=\pm1$ spin states, allowing us to define a qubit subspace consisting of the $m_s=0$ and $m_s=-1$ states. Microwave pulses for qubit control are delivered via a gold stripline, which is embedded into the sample mirror \cite{bogdanovic_robust_2017}.

All reported uncertainties in this work correspond to one standard deviation confidence intervals.

\begin{figure*}[ht]
    \includegraphics[scale=0.8]{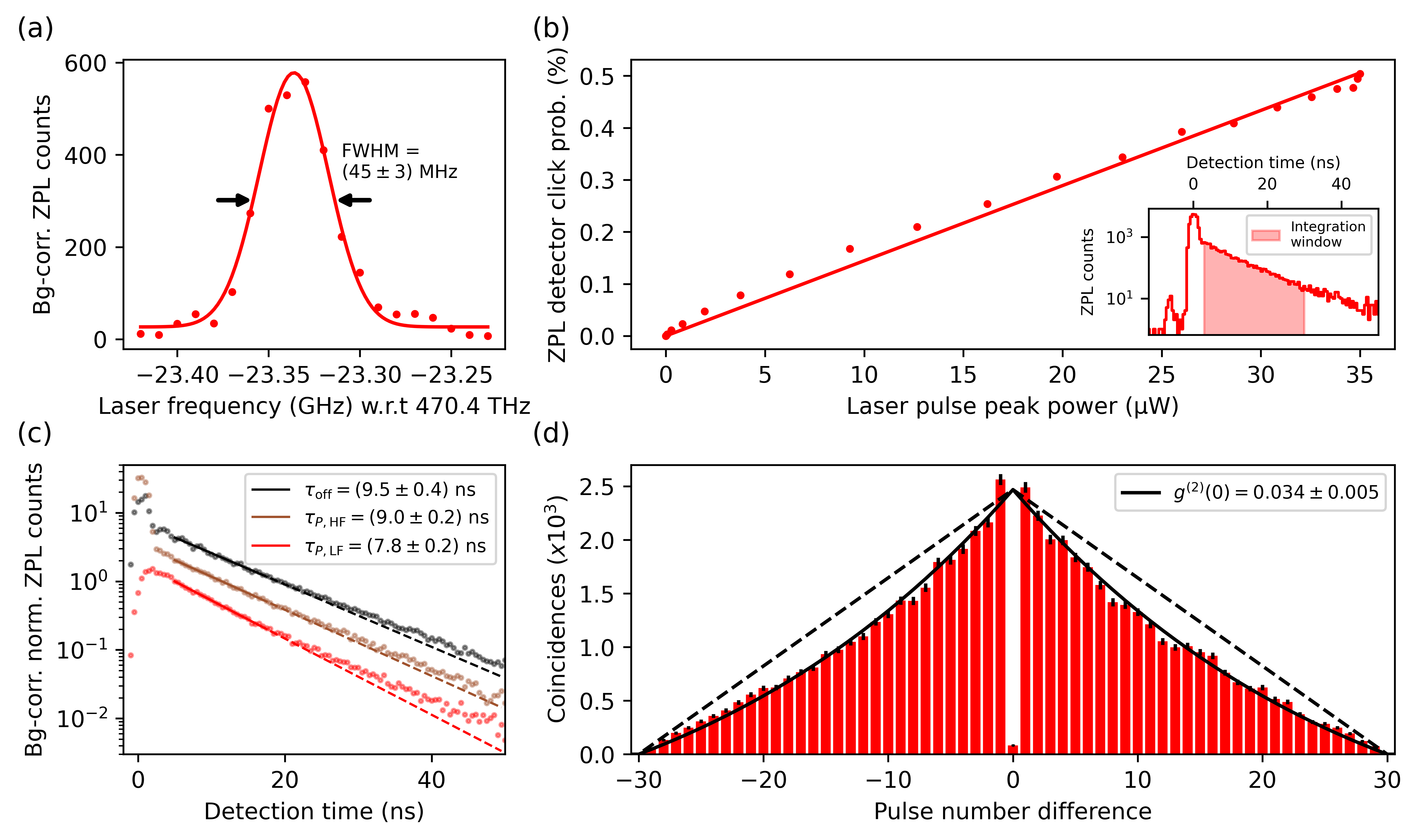}
    \caption{Characterization of the optical interface of the cavity-coupled NV center. (a) PLE measurement of the NV center's $E_y$ transition (readout transition) in the LF cavity mode with a Gaussian fit (solid line). Details about the PLE measurement sequence and the applied background correction are presented in Appendix~\ref{apx:PLEs}. (b) Pulsed resonant saturation measurement of the readout transition in the LF cavity mode, together with a linear fit as a guide for the eye. The inset shows the time-resolved detector counts for a laser pulse peak power of $\unit[35]{\upmu W}$ measured free-space before the objective and the used integration window from $\unit[3]{ns}$ to $\unit[30]{ns}$ with respect to the excitation pulse center. The small peak at a detection time of about $\unit[-5]{ns}$ is light of the excitation pulse that is backscattered into the free-space detection before reaching the cryostat. (c) Lifetime measurements of the NV center's $E_y$ excited state on cavity resonance in the LF and HF cavity modes, as well as off resonance in the LF mode. Details about the applied background correction are outlined in Appendix~\ref{apx:lifetime}. The fit windows of $\unit[5]{ns}$ to $\unit[18]{ns}$ are represented by the length of the solid lines of the monoexponential fits. The data is offset for visual clarity. (d) Second-order correlation measurement of the readout transition in the LF cavity mode for a pulse train of $30$ consecutive short resonant excitation pulses. In this measurement, the integration window of (b) is used. The triangular function capturing the finite pulse train (black dashed line) is shown next to the fit function that also includes the spin flipping process (black solid line).}
    \label{fig:NV_optical_characterization}
\end{figure*}

\section{Coupling a single Nitrogen-Vacancy center to the Microcavity}
With the full spatial and spectral tunability of the microcavity \cite{bogdanovic_design_2017}, we select a position on the diamond membrane that is close to the mirror-embedded gold stripline, exhibits a high cavity quality factor, and a well-coupled NV center. At the selected position, a frequency splitting between the horizontal and vertical polarization cavity mode of $\unit[(9.56\pm0.02)]{GHz}$ is observed, which we attribute to birefringence due to the presence of strain in the diamond membrane. The individual cavity polarization mode shows a linewidth of $\unit[(1.69\pm0.02)]{GHz}$, which corresponds to a quality factor of $280 \times 10^3$. With the determined cavity geometry, a cavity mode volume of $\unit[86]{\lambda^3}$ is simulated. Accounting for our residual cavity vibration level of $\unit[22]{pm}$, this results in a maximal vibration-averaged Purcell factor of $12$ for a NV center that is perfectly coupled to a single cavity mode \cite{herrmann_coherent_2024}. In addition, a cavity mode outcoupling efficiency through the sample mirror of $\unit[39]{\%}$ is calculated by taking the mirror coating transmission losses into account (see Appendix~\ref{apx:cavity_characterization} for the cavity characterization details).

For the investigated cavity we find a NV center that is coupled with its $E_y$ transition (readout transition) at a frequency of $\unit[470.377]{THz}$ ($\unit[-23.34]{GHz}$ with respect to $\unit[470.4]{THz}$). Figure~\ref{fig:NV_optical_characterization}(a) shows a PLE measurement of this transition in the LF cavity mode, resulting in a linewidth of $\unit[(45\pm3)]{MHz}$ determined by a Gaussian fit. We attribute the additional broadening above the expected linewidth of $1/2\pi\tau_{P,\text{LF}}\approx\unit[20]{MHz}$ for the Purcell-reduced lifetime $\tau_{P,\text{LF}}$ in the LF cavity mode as stated below to spectral diffusion. The determined transition linewidth is typical for NV centers with bulk-like optical properties, and the used experimental PLE sequence involving $\unit[515]{nm}$ repump laser pulses  \cite{ruf_optically_2019}. Furthermore, we find the NV center's $E_x$ transition at a frequency of $\unit[10]{GHz}$. This corresponds to a lateral strain with $E_x~\text{-}~E_y$ excited state splitting of $\sim \unit[33]{GHz}$, which is larger compared to typical bulk samples. We note that we also find less strained NV centers in the diamond membrane (see Appendix~\ref{apx:second_nv} for measurements of a second cavity-coupled NV center). Next to the NV center transitions associated with the $m_s=0$ spin state, we find emission lines at $\unit[-26.0]{GHz}$ as well as $\unit[-24.5]{GHz}$ that are associated with the $m_s=\pm1$ spin states. We attribute these to the NV center's $E_1$ and $E_2$ transitions, where we use the former as our spin initialization transition. In all measurements, a static magnetic field of about $\unit[37.5]{G}$ is present along the NV center crystal axis. See Appendix~\ref{apx:PLEs} for the PLE measurements of the other transitions and details about the used sequences.

In the sequences of all following measurements, we start by applying a $\unit[515]{nm}$ charge repump laser pulse to prepare the NV center with high probability in its negatively charged state, and a spin initialization laser pulse to initialize the NV center in its $m_s=0$ spin state (see Appendix~\ref{apx:init_and_readout} for details). Figure~\ref{fig:NV_optical_characterization}(c) shows lifetime measurements of the NV center's $E_y$ excited state using $\unit[2]{ns}$ short resonant excitation pulses (see Appendix~\ref{apx:excitation_pulse} for details about the excitation pulse). The Purcell-reduced lifetime is measured in the LF and HF cavity modes with the cavity being on resonance with the readout transition. An off resonance lifetime is determined in the LF mode by detuning the cavity from the readout transition by $\unit[-4]{GHz}$. The lifetime measurement data is fitted with monoexponential curves. We observe that the data start to deviate from a monoexponential behavior for detection times $\gtrsim\unit[20]{ns}$. We attribute this to signal of other (weaker coupled) NV centers or fluorescence of the cavity mirrors, and therefore restrict the fit window size. Details about the applied background subtraction and the influence of the fit window are further outlined in Appendix~\ref{apx:lifetime}.
From the monoexponential fit, we determine an off resonance lifetime of $\tau_\text{off}=\unit[(9.5\pm0.4)]{ns}$ and a Purcell-reduced lifetime of $\tau_{P,\text{LF}}=\unit[(7.8\pm0.2)]{ns}$ and $\tau_{P,\text{HF}}=\unit[(9.0\pm0.2)]{ns}$ for the LF and HF cavity mode, respectively. The off resonance lifetime differs from the expected excited state lifetime of $\unit[12.4]{ns}$ for low-strained NV centers \cite{hermans_entangling_2023}. We attribute this to strain-induced mixing in the excited states \cite{manson_nitrogen-vacancy_2006,goldman_state-selective_2015}. In Appendix~\ref{apx:lifetime} we show a complementary investigation of this mechanism with time-resolved detection during continuous wave resonant excitation. The determined lifetimes allow to calculate the Purcell factor
\begin{equation}
    F_P = \frac{1}{\beta_0}\left(\frac{\tau_\text{off}}{\tau_P} - 1\right),
\label{equ:purcell_factor}
\end{equation}
with the NV center Debye-Waller factor $\beta_0=0.03$ \cite{riedel_deterministic_2017}.
The Purcell factors for the LF and HF cavity mode are $F_{P,\text{LF}}=\unit[7.3\pm1.6]{}$ and $F_{P,\text{HF}}=\unit[2.0\pm1.4]{}$. In all the following measurements, we use the stronger coupling to the LF cavity mode to enhance the NV center emission and the HF cavity mode for excitation. Based on the determined Purcell factor the NV center coherently emits $\beta_\text{LF} = \beta_0 F_{P,\text{LF}}/(\beta_0 F_{P,\text{LF}} + 1)\approx\unit[18]{\%}$ into the LF cavity mode, which corresponds to a cavity outcoupled ZPL emission of about $\unit[7]{\%}$ and a maximal detector click probability of about $\unit[1.4]{\%}$ per excitation pulse for our setup efficiency and detection time window (see Appendix~\ref{apx:parameters} for details).

In Figure~\ref{fig:NV_optical_characterization}(b), the saturation behavior of the NV center readout transition is studied. For that, the peak power of the short excitation laser pulse is varied, and the ZPL detector click probability after the first excitation pulse is measured. A ZPL detector click probability of $\unit[0.5]{\%}$ per pulse is determined for the highest power applied, which outperforms the standard NV center quantum network node setups based on solid immersion lenses by an order of magnitude \cite{hensen_loophole-free_2015,humphreys_deterministic_2018,pompili_realization_2021,hermans_qubit_2022}. In addition, the detector click probability still increases in a linear fashion for the investigated laser power regime showing that higher detector click probabilities can be reached if more laser power is applied. In the experiments, we are limited to the investigated laser power regime due to the efficiency of our short optical laser pulse generation setup.

Furthermore, we confirm single-photon emission by performing a pulsed resonant HBT experiment on the readout transition. In the experiment, we perform in total $55\times10^6$ measurement sequence repetitions in which we apply a train of short excitation pulses with a consecutive time separation of $\unit[126]{ns}$. Figure~\ref{fig:NV_optical_characterization}(d) shows the second-order correlation function for $30$ consecutive excitation pulses. The second-order correlation function exhibits clear antibunching at zero pulse number difference. Further, it shows two bunching-like features, which we attribute to the spin flipping probability into the $m_s=\pm1$ states for larger pulse number differences and the probability to decay back into the $m_s=0$ state from the intersystem crossing singlet state for small pulse number differences. We fit the data with a triangular function as expected for the finite pulse train multiplied by a monoexponential function capturing the spin flipping process. The exponential decay for small pulse number differences is not included, since the amount of affected data points is insufficient for a reliable fit. We quantify an antibunching value of $g^{(2)}(0)=\unit[0.034\pm0.005]{}$ by the ratio of measured coincidences in the zero bin to the triangular fit amplitude.

\begin{figure*}[ht]
    \includegraphics[scale=0.8]{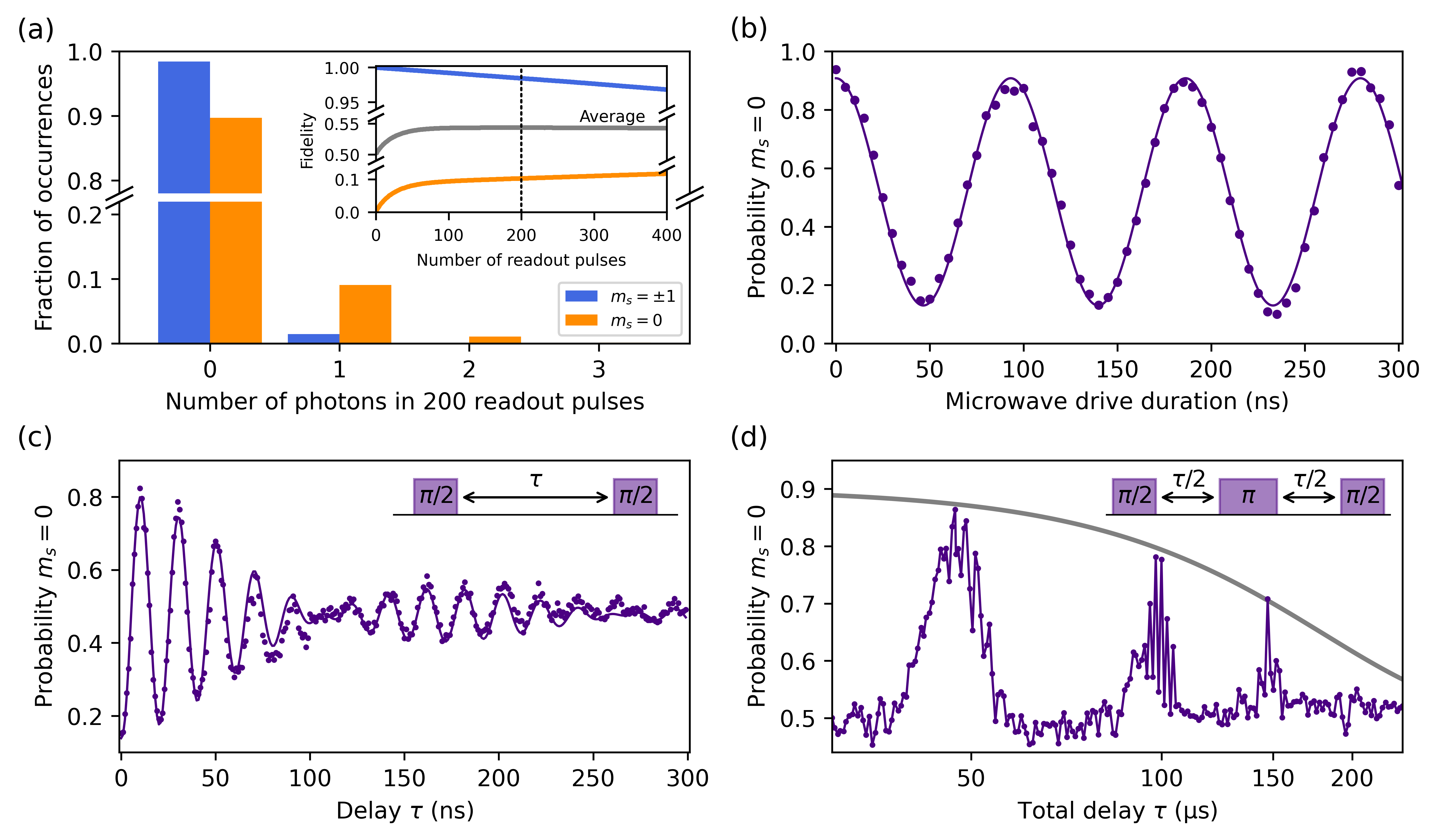}
    \caption{Characterization of the NV center electron spin qubit. (a) Statistics of detected photons for our spin qubit readout after spin initialization in the $m_s=\pm1$ or $m_s=0$ state. The inset presents the corresponding readout fidelities depending on the number of short excitation pulses used. For the experiments $200$ short readout pulses are used, which leads to fidelities of $F_{\pm1} = \unit[98.4]{\%}$, $F_0 = \unit[10.3]{\%}$ and $F_\text{avg} = \unit[54.4]{\%}$ for the shown calibration with $500 \times 10^3$ repetitions. (b) Coherent Rabi oscillations with a fit (solid line) showing a Rabi frequency of $\Omega=\unit[(10.73\pm0.02)]{MHz}$. The error bars are within the dot size. (c) Ramsey measurement with an artificial detuning of $\unit[50]{MHz}$. The inset shows the used microwave pulse sequence, where a delay time-dependent phase is applied to the second pulse to generate the artificial detuning. The fit (solid line) captures a beating signature of $\unit[(2.68\pm0.05)]{MHz}$ which is attributed to the coupled nitrogen nuclear spin and the free-induction decay of $T_2^*=\unit[(170\pm20)]{ns}$ with an exponent of $n=1.0\pm0.2$. The error bars are within the dot size. (d) Hahn-Echo measurement with its microwave pulse sequence in the inset. The grey solid line is a guide for the eye of a Gaussian decay curve with an amplitude of $0.9$ and a decay time of $\unit[180]{\upmu s}$. The error bars are within the dot size.}
    \label{fig:NV_spin_characterization}
\end{figure*}

\section{Coherent Microwave Control of the Nitrogen-Vacancy center spin qubit}
The NV center electron spin is coherently manipulated with microwave pulses, which are delivered via an about $\unit[10]{\upmu m}$ distant gold stripline. Before microwave spin manipulation a $\unit[515]{nm}$ charge repump laser pulse prepares the NV center with high probability in its negatively charged state and the spin is optically initialized in its $m_s=0$ ground state with a spin initialization laser pulse and a fidelity of $\unit[(93.5\pm0.9)]{\%}$ (see Appendix~\ref{apx:init_and_readout} for details about the initialization and qubit readout analysis). The applied magnetic field along the NV center crystal axis splits the $m_s=\pm1$ states by about $\unit[210]{MHz}$, enabling selective driving of the $m_s=0$ and $m_s=-1$ qubit subsystem. After spin manipulation, the qubit is read out via the readout transition using the short excitation pulses and the same integration window as introduced in Fig.~\ref{fig:NV_optical_characterization}(b). Figure~\ref{fig:NV_spin_characterization}(a) shows the statistics of detected photons for our spin qubit readout after spin initialization in the $m_s=\pm1$ or $m_s=0$ state. This measurement is used as a readout fidelity calibration, and its dependency on the number of used short readout pulses is presented in the inset. The average fidelity $F_\text{avg}$ of the $m_s=\pm1$ readout fidelity $F_{\pm1}$ and the $m_s=0$ readout fidelity $F_0$ plateaus for larger readout pulse numbers and is optimal around $200$ pulses, which is used in the following experiments. These readout fidelities are used for qubit readout correction, where the finite $m_s=0$ spin state initialization fidelity is also taken into account. We note that the average readout fidelity is not limited by laser power, but rather the average number of emitted photons before the studied NV center spin flips. The outcoupled ZPL emission of about $\unit[7]{\%}$ for our system is comparable to the collected non-coherent phonon sideband (PSB) emission used for readout in a SIL setup, rendering high-fidelity qubit readout possible \cite{robledo_high-fidelity_2011}.

In Figure~\ref{fig:NV_spin_characterization}(b), coherent Rabi oscillations with a Rabi frequency of $\Omega=\unit[(10.73\pm0.02)]{MHz}$ are shown for our spin qubit at a microwave frequency of $\unit[2.773]{GHz}$. Based on the fitted contrast of the Rabi oscillations and correcting for the finite $m_s=0$ spin state initialization fidelity, a microwave $\pi$ pulse fidelity of $\unit[(83\pm9)]{\%}$ is extracted.

For the next experiments, a $\pi$ pulse and a $\pi/2$ pulse are calibrated by varying the amplitude of five consecutively applied $\pi$ pulses that minimize the probability to readout $m_s=0$. The $\pi$ pulse duration is fixed to $\unit[64]{ns}$, and half of that duration is used for the $\pi/2$ pulse. As depicted in Fig.~\ref{fig:NV_spin_characterization}(c), the qubit coherence is probed in a Ramsey experiment with an artificial detuning of $\unit[50]{MHz}$ by applying a delay time-dependent phase shift to the second $\pi/2$ pulse. The observed beating signature is attributed to the hyperfine coupling of the nitrogen nuclear spin, which is included in the fit model (see Appendix~\ref{apx:ramsey_analysis} for details). By fitting the measurement data, a free-induction decay time of $T_2^*=\unit[(170\pm20)]{ns}$ is determined, which is short compared to typical $T_2^*$ times for these diamond devices (see Appendix~\ref{apx:second_nv} for data of a second cavity-coupled NV center).

Using a Hahn-Echo experiment as presented in Fig.~\ref{fig:NV_spin_characterization}(d), the qubit coherence can be largely recovered by choosing an appropriate total delay time $\tau$, which mitigates quasi-static magnetic field fluctuations. In this measurement, three revivals are observed and a Gaussian decay curve with a time constant of $\unit[180]{\upmu s}$ is displayed as a guide for the eye in Fig.~\ref{fig:NV_spin_characterization}(d). The revivals occur around decoupling interpulse delay times $\tau/2$ that correspond to integer multiples of the bare Larmor period of the carbon-13 nuclear spin bath of about $\unit[25]{\upmu s}$ ($\gamma_{C13}=\unit[1.0705]{kHz/G}$ and $B=\unit[37.5]{G}$) \cite{ryan_robust_2010}.

\begin{center}
\begin{figure}[ht]
    \includegraphics[width=\linewidth]{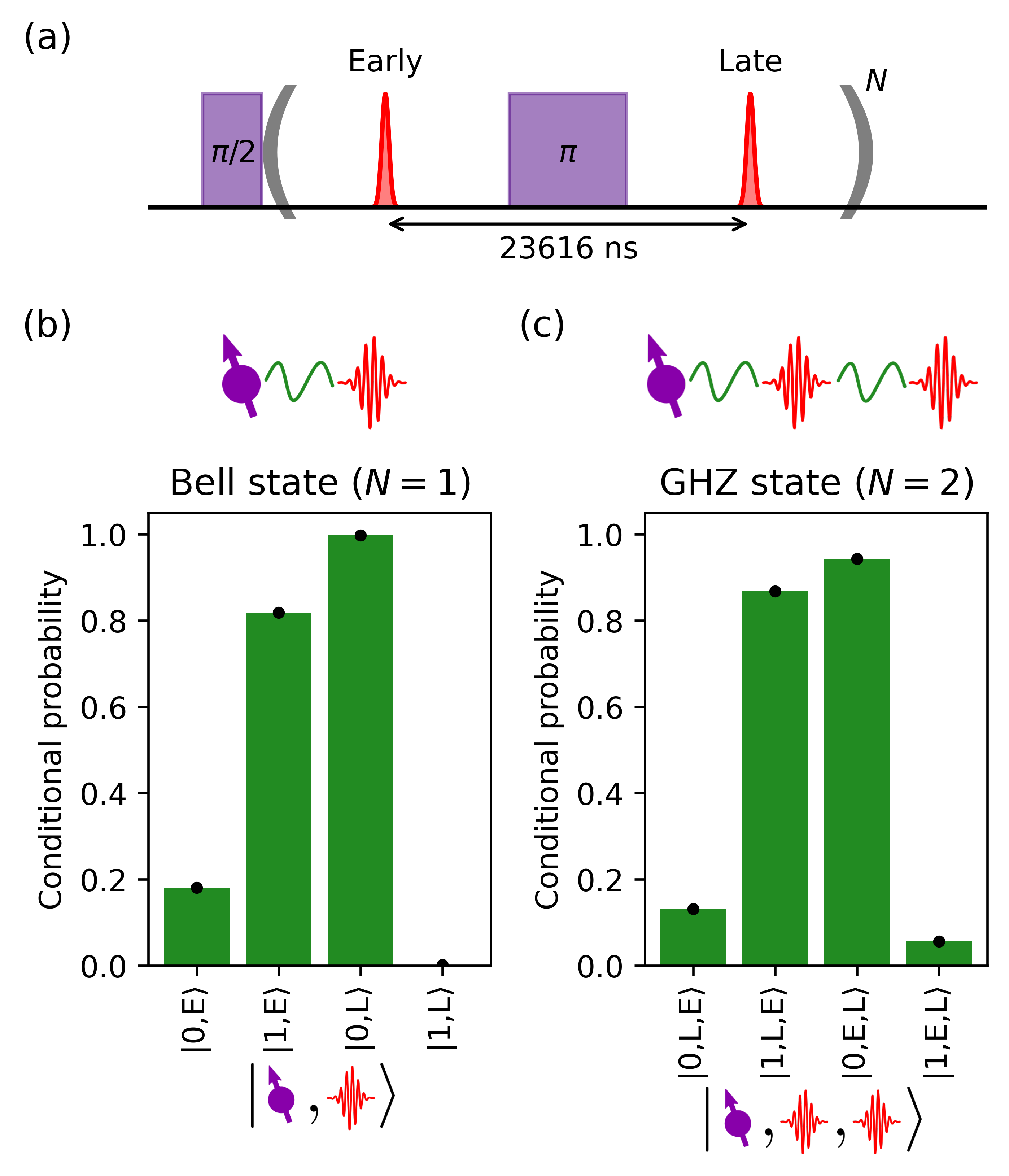}
    \caption{Spin-photon correlation measurements. (a) Pulse sequence to generate spin-photon states involving microwave pulses (in purple) for qubit control and short excitation pulses (in red) to generate spin-dependent time-bin photons. In the experiments, a microwave decoupling interpulse delay time of $\tau/2 = \unit[23568]{ns}$ is used. (b) Conditional probabilities of the spin qubit readout in the Z-basis after successful heralding an early ($\ket{\text{E}}$) or late ($\ket{\text{L}}$) photon of the spin-photon Bell state. The error bars are within the black dot size. (c) Conditional probabilities of the spin qubit readout in the Z-basis after a successful double heralding event of the photonic state $\ket{\text{L},\text{E}}$ or $\ket{\text{E},\text{L}}$ of the spin-photon GHZ state. The error bars are within the black dot size.}
    \label{fig:spin_photon_correlations}
\end{figure}
\end{center}

\section{Generation of Spin-Photon states}
In this section, we use our system to generate two-qubit and three-qubit spin-photon states and measure heralded correlations between the photonic and spin states. Figure~\ref{fig:spin_photon_correlations}(a) depicts the used pulse sequence, which combines spin qubit microwave control with the short resonant excitation pulses to generate spin-dependent photonic time-bin qubits. After optical initialization, the spin qubit is brought into equal superposition with a $\pi/2$ microwave pulse, followed by two resonant excitation pulses with an intermediate microwave $\pi$ pulse. This generates the Bell state $\ket{\text{NV spin},\text{photon}} = (\ket{1,\text{E}} + \ket{0,\text{L}})/\sqrt{2}$ between the spin qubit states $m_s=0$ ($\ket{0}$) and $m_s=-1$ ($\ket{1}$) and the photonic qubit in the time-bin basis of early ($\ket{\text{E}}$) and late ($\ket{\text{L}}$) as used for remote entanglement generation in quantum network demonstrations \cite{bernien_heralded_2013,pfaff_unconditional_2014,hensen_loophole-free_2015}. The generated photon is measured by a single-photon detector, which heralds an early or late detection event. The spin qubit is optically read out in its Z-basis by $200$ short readout pulses. The qubit readout probabilities conditioned on a photon heralding event in the early or late time-bin are shown in Fig.~\ref{fig:spin_photon_correlations}(b) (see Appendix \ref{apx:init_and_readout} for details about the applied qubit readout correction and Appendix \ref{apx:spin_photon_correlations_x_basis} for an X-basis spin qubit readout). A photon detection event in the late time-bin heralds the qubit in its expected $\ket{0}$ state. For an early time-bin photon detection event, the spin is read out with a probability of $\unit[82]{\%}$ in its expected $\ket{1}$ state. We attribute the lower readout probability of the $\ket{1}$ state to our limited microwave $\pi$ pulse fidelity and an imperfect pulse calibration. In the Bell state correlation measurement, we record $27143$ photon heralding events in $5\times10^6$ attempts, which corresponds to a probability of $\unit[0.54]{\%}$ per attempt.

Furthermore, the Greenberger–Horne–Zeilinger (GHZ) state $\ket{\text{NV spin},\text{photon},\text{photon}} = (\ket{0,\text{E},\text{L}} + \ket{1,\text{L},\text{E}})/\sqrt{2}$ can be generated by repeating the center sequence with the two short resonant excitation pulses before readout (see case $N=2$ in Fig.~\ref{fig:spin_photon_correlations}(a)). The correlations after a successful double photon heralding event are displayed in Fig.~\ref{fig:spin_photon_correlations}(c). After heralding a correct two photon state, the expected spin state is read out with similar probabilities as in the Bell state measurement. In total, we measure $5242$ two photon heralding events in $220\times10^6$ attempts, which corresponds to a probability of $2.4\times10^{-5}$ per attempt.

\section{Conclusion}
We have equipped a coherent NV center spin qubit with an efficient spin-photon interface by coupling it to an open microcavity. Coherent qubit control is realized with microwave pulses, and qubit initialization as well as qubit readout are achieved with optical pulses in a cross-polarized resonant excitation and detection scheme. Moreover, we have demonstrated the quantum networking capabilities of our system by generating two-qubit and three-qubit spin-photon states and measuring heralded correlations between the photonic time-bin and the spin qubit states.

For the presented system, we project a saturation ZPL detector click probability of $\unit[1.4]{\%}$ per pulse, by improving the excitation efficiency without modifying the microcavity. Moreover, the system can be optimized by implementing a charge-resonant check procedure of the optical transitions to remove qubit initialization infidelities \cite{brevoord_heralded_2024} and microwave pulse shaping for high-fidelity qubit control \cite{bartling_universal_2025}.

Our work opens up opportunities to explore quantum networking with Purcell-enhanced NV centers, promising remote entanglement generation with higher rates and fidelities. Furthermore, we expect that the here developed methods provide guidance for other solid-state qubits to realize efficient spin-photon interfaces based on microcavities for quantum network applications \cite{bayer_optical_2023,zifkin_lifetime_2024,herrmann_coherent_2024,hessenauer_cavity_2025,ulanowski_spectral_2024}.

\begin{acknowledgments}
We thank Raymond Vermeulen for designing and building the standalone pulse generator used to modulate the electro-optical amplitude modulator. We thank Martin Eschen for laser-ablating the fiber mirror. We thank Jiwon Yun, Kai-Niklas Schymik, Conor Bradley, Alexander Stramma, and Mariagrazia Iuliano for helpful discussions. We thank Alexander Stramma for feedback on the manuscript.

We acknowledge financial support from the Dutch Research Council (NWO) through the Spinoza prize 2019 (project number SPI 63-264) and from the EU Flagship on Quantum Technologies through the project Quantum Internet Alliance (EU~Horizon~2020, grant agreement no.~820445).

\textbf{Author contributions:} J.F. and Y.H. contributed equally to this work. J.F. and Y.H. conducted the experiments and analyzed the data. J.F., Y.H., and S.S. developed and built the cross-polarization filtering setup. J.F. developed the cavity lock and the polarization suppression control. J.F., Y.H., and C.F.J.W. developed and built the short optical laser pulse generation setup. J.F. and C.F.J.W. developed the microwave qubit control. S.S. characterized the cavity fiber. M.R. fabricated the diamond device. J.F., Y.H., and R.H. wrote the manuscript with input from all authors. R.H. supervised the experiments.

\textbf{Data availability:} The datasets that support this manuscript are available at 4TU.ResearchData \cite{fischer_data_2025}.

\end{acknowledgments}

\appendix
\newpage
\section{Experimental setup\label{apx:setup}}
We use a floating stage closed-cycle optical cryostat (Montana~Instruments~HILA) with a base temperature of about $\unit[6]{K}$ and a sample mirror holder temperature of about $\unit[8]{K}$. The spherical fiber mirror, which faces the sample mirror, is made out of a single-mode optical fiber with a pure silica core and a polyimide protection coating (Coherent FUD-4519,~S630-P) and is mounted on a cryo-compatible positioning stage (JPE CPSHR1-a). A room temperature objective (Zeiss LD~EC~Epiplan-Neofluar), which is thermally shielded, is used to optically interface the sample mirror side of the cavity. The objective is positioned with three linear nanopositioning stages (Physik~Instrumente Q-545) in a tripod configuration. A permanent neodymium disc magnet (Supermagnete S-70-35-N), which is mounted on the lid of the cryostat, is used to create a static magnetic field at the position of the sample mirror. Further details about the cryogenic setup can be found in Ref.~\cite{herrmann_low-temperature_2024}.

The setup is operated with a computer and the Python-based software QMI version 0.44 \cite{raa_qmi_2023}. The data is stored and analyzed with the Python framework quantify-core.
The actual measurement sequences are timed and executed by an interplay between a microcontroller (J\"{a}ger~Computergesteuerte~Messtechnik Adwin~Pro~II) and an arbitrary waveform generator (Tektronix AWG5014C). Microwave pulses are generated with a single sideband modulated vector signal generator (Rohde~\&~Schwarz SMBV100A), amplified by a microwave amplifier (Mini-Circuits ZHL-50W-63+), and delivered to the cryostat via a home-built microwave switch. At the qubit frequency of $\unit[2.773]{GHz}$ we measure about $\unit[26]{dB}$ transmission loss through the cryostat. Further details about the microwave wiring and a picture of the sample mirror with mirror-embedded gold striplines can be found in Ref.~\cite{herrmann_low-temperature_2024}.

\subsection{Laser pulse generation and delivery}
We use one $\unit[637]{nm}$ continuous wave diode laser each for cavity locking (Newport Newfocus~Velocity~TLB-6300-LN) and spin initialization (Toptica DL~Pro~$\unit[637]{nm}$). To modulate the laser intensity and generate microsecond-long optical pulses with high on/off ratios, two cascaded in-fiber acousto-optic modulators (AOM, Gooch~and~Housego Fiber-Q~$\unit[633]{nm}$) are used. Both lasers are combined free-space by a 50:50 non-polarizing beamsplitter (Thorlabs BS016). Then, a Glan-Taylor polarizer (Thorlabs GT10-A) and a half-wave and quarter-wave plate (Thorlabs WPH05M-633 and WPQ05M-633) are used to set and control the polarization. Finally, the $\unit[637]{nm}$ lasers are overlapped with the $\unit[515]{nm}$ repump laser (H\"{u}bner~Photonics Cobolt~06-MLD) by a dichroic mirror (Semrock~Di01-R532) and coupled into a single-mode fiber, which is connected to the cavity fiber.

A frequency-doubled, high-power tunable diode laser (Toptica TA-SHG~Pro~$\unit[637]{nm}$) is used for readout as well as for cross-polarization control. For that, the laser is split by a 50:50 fiber beamsplitter (Thorlabs PN635R5A2) into two in-fiber paths. In the path for cross-polarization control, two in-fiber AOMs are used for modulation, whereas in the readout laser path, a temperature-stabilized electro-optical amplitude modulator (EOM, Jenoptik AM635) is deployed additionally. This allows for faster intensity modulation and the generation of nanosecond-short resonant excitation pulses. The EOM is DC biased by a programmable bench power supply (Tenma 72-13360) using a bias-tee (Mini-Circuits ZX85-12G-S+) and a DC block (Mini-Circuits BLK-89-S+) to control a constant transmission level. In addition, fast electrical pulses are generated with a home-built standalone pulse generator~\cite{vermeulen_pulse_2020} and applied to the EOM via the RF input of the bias-tee to generate nanosecond-short excitation pulses (see Appendix \ref{apx:excitation_pulse} for their characterization). Finally, both in-fiber paths are combined with a 75:25 fiber beamsplitter (Thorlabs PN635R3A2) whose $\unit[75]{\%}$ output is connected to a polarization-maintaining fiber (Thorlabs P3-630PM-FC-2) before free-space launching. The $\unit[25]{\%}$ output is connected to a power meter (Thorlabs PM100USB with head S130VC) for laser power monitoring. To preserve a high polarization extinction ratio in both paths in-line fiber polarizers (Thorlabs ILP630PM-APC) are used.

All $\unit[637]{nm}$ lasers are stabilized on a wave meter (HighFinesse WS-U).

\subsection{Free-space cross-polarization optics}
The cavity is interfaced from the sample mirror side with free-space cross-polarization optics as depicted in Fig.~\ref{fig:microcavity_schematic}(b). The collimated beam leaving the cryostat chamber is guided by broadband dielectric mirrors (Thorlabs BB03-E03) on the optical table and is reflected off a dichroic mirror (Semrock Di02-R635) before reaching the polarization optics. Next, the beam passes through half-wave and quarter-wave plates (Thorlabs WPH05M-633 and WPQ05M-633), which are mounted in piezo rotation stages (Newport AG-PR100 controlled by Newport AG-UC8). These wave plates map the excitation and detection polarization of the polarizing beamsplitter cube (PBS, Thorlabs PBS202) to the cavity polarization modes.

In the excitation path of the PBS the readout laser is launched from the polarization-maintaining fiber into free-space (Thorlabs KT120 with objective RMS10X) and passes a Glan-Thompson polarizer (Thorlabs GTH10M-A), followed by a quarter-wave and half-wave plate, which are mounted in piezo rotation stages as well.

In the detection path of the PBS, the light is filtered by a Glan-Thompson polarizer (Thorlabs GTH10M-A), an angle-tunable etalon (Light~Machinery custom coating, $\approx \unit[100]{GHz}$ full width at half maximum (FWHM) at $\unit[637]{nm}$), and a bandpass filter (Thorlabs FBH640-10). Finally, the light is coupled (Thorlabs KT120 with objective RMS10X) into a single-mode fiber (Thorlabs P3-630A-FC-2) and detected by a single-photon detector (Picoquant Tau-SPAD-20). For the HBT experiment shown in Fig.~\ref{fig:NV_optical_characterization}~(d), a 50:50 fiber beamsplitter (Thorlabs TW630R5A2) is added, and a second single-photon detector (Laser~Components Count-10C-FC) is used. The detectors are connected to a single-photon counting module (Picoquant~Hydraharp~400) and the counter module of the microcontroller. The gating of the single-photon detectors is used to protect them against blinding during the application of the spin initialization laser.

\subsection{Microcavity operation}
For the resonant cross-polarization excitation and detection scheme used in this work, it is essential to keep the cavity on resonance with the NV center and to maintain a high excitation laser suppression during the measurements. To ensure these operational conditions, we probe the cavity resonance as well as the excitation laser suppression interleaved with the experimental sequences and stream the data to a computer in real-time to run an optimization routine next to the measurement. In both cases, the probe laser light is detected with the ZPL single-photon detector and recorded with the counter module of the microcontroller. This technique allows for live feedback on the order of Hertz, which is sufficient to compensate for drifts of the cavity resonance and the cross-polarization filtering. With the methods outlined in the following, it is possible to run measurements remotely for days.

To keep the cavity on resonance with the readout transition of the NV center, a side-of-fringe lock is deployed. For that, the cavity lock laser is frequency-tuned to the fringe of the cavity mode, and the single-photon detector counts are recorded during interleaved $\unit[100]{\upmu s}$ long probe laser pulses. The recorded counts are streamed to a computer that uses a proportional control loop programmed in Python to optimize on a specified set value. The proportional control loop feeds back on the piezo that is controlling the fiber mirror position and, by that, locks the cavity on resonance. The cavity lock laser power is adjusted such that a count rate of about $\unit[700]{kHz}$ is measured when the cavity is on resonance with the cavity lock laser. For every specified set value, the cavity lock laser frequency can be adjusted such that the cavity is on resonance with the NV center.

For the cross-polarization filtering, interleaved $\unit[10]{nW}$ probe pulses of the cross-polarization control laser are applied for $\unit[400]{\upmu s}$ while recording the single-photon detector counts. These counts represent the current excitation laser suppression and are streamed to a computer that runs a hill climbing type algorithm programmed in Python. The algorithm makes iterative changes to the four piezo rotation mounted wave plates as introduced in Fig.~\ref{fig:microcavity_schematic}(b) to minimize the recorded counts. With this polarization suppression control loop, typical long-term count rates are $\unit[<2]{kHz}$ for $\unit[10]{nW}$ probe pulses, which corresponds to excitation laser suppression values $\unit[>60]{dB}$.

\section{Hybrid cavity characterization\label{apx:cavity_characterization}}
\begin{center}
\begin{figure}[ht]
    \includegraphics[width=\linewidth]{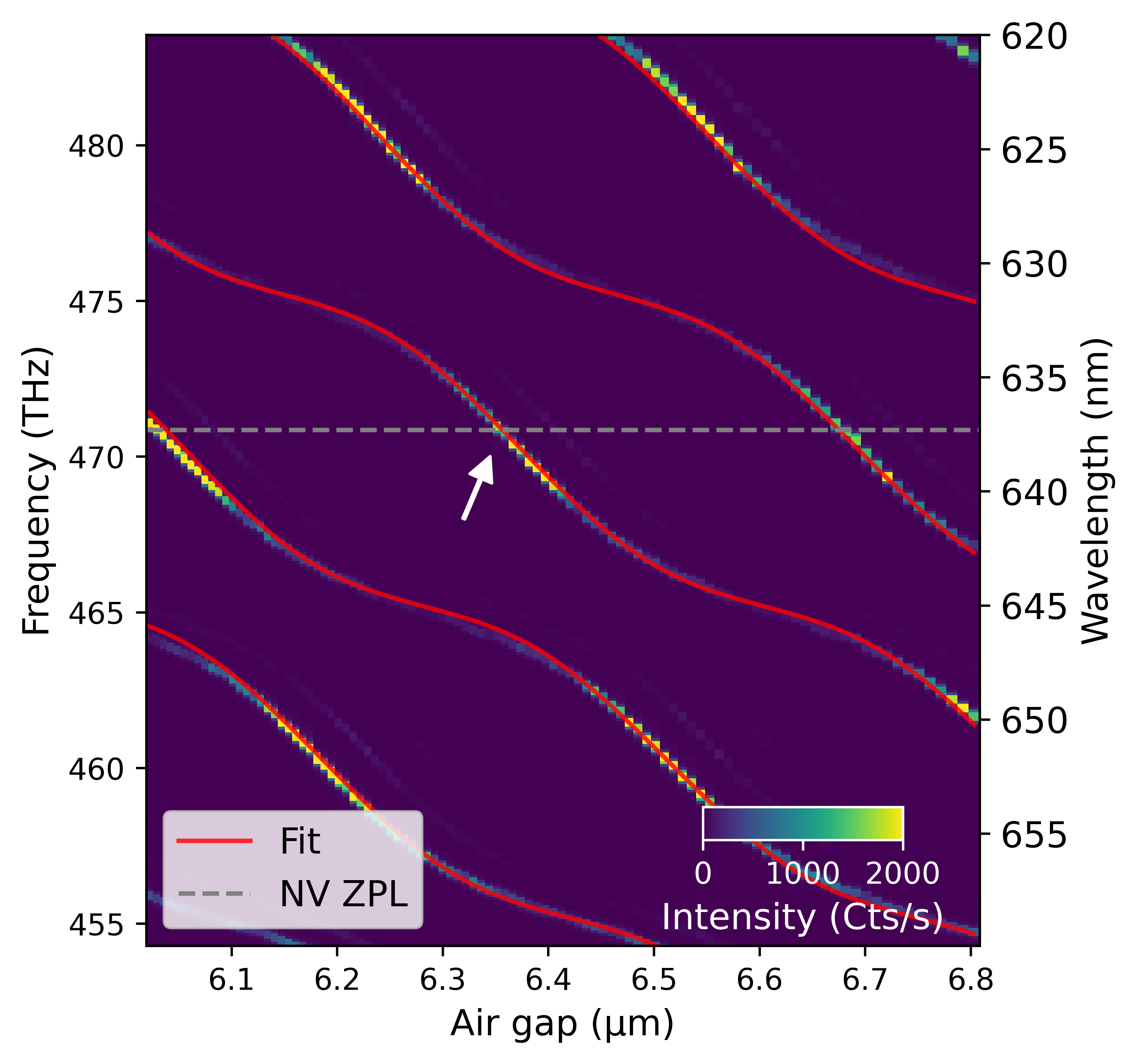}
    \caption{Cavity mode dispersion measurement of the hybrid air-diamond cavity. The cavity transmission of a broadband white light source spanning from $\unit[600]{nm}$ to $\unit[700]{nm}$ is measured on a spectrometer depending on the air gap. The bright fundamental modes are extracted and fitted by an analytical equation \cite{janitz_fabry-perot_2015} to determine the air gap and the diamond thickness. In all measurements, the cavity is operated in the air-like mode at $\unit[637]{nm}$ as indicated by the white arrow.}
    \label{fig:cavity_dispersion}
\end{figure}
\end{center}

All measurements presented in the main text are performed at the same cavity length and lateral cavity position. By measuring the cavity mode dispersion as shown in Fig.~\ref{fig:cavity_dispersion} we extract an air gap of $\unit[6.39]{\upmu m}$ and a diamond thickness of $\unit[6.20]{\upmu m}$, which corresponds to an air-like cavity mode \cite{janitz_fabry-perot_2015} with hybrid cavity mode number $67$. Due to a non-uniform bond, there might be a bond gap between the sample mirror and the diamond membrane. Further, a cavity mode dispersion slope of $\unit[34]{MHz/pm}$ and an effective cavity length of $\unit[13.2]{\upmu m}$ is estimated \cite{van_dam_optimal_2018}. Together with the $\unit[21.4]{\upmu m}$ radius of curvature of the fiber mirror, a cavity beam waist of $\unit[1.46]{\upmu m}$ and a cavity mode volume of $\unit[86]{\lambda^3}$ is calculated.

\begin{center}
\begin{figure}[ht]
    \includegraphics[width=\linewidth]{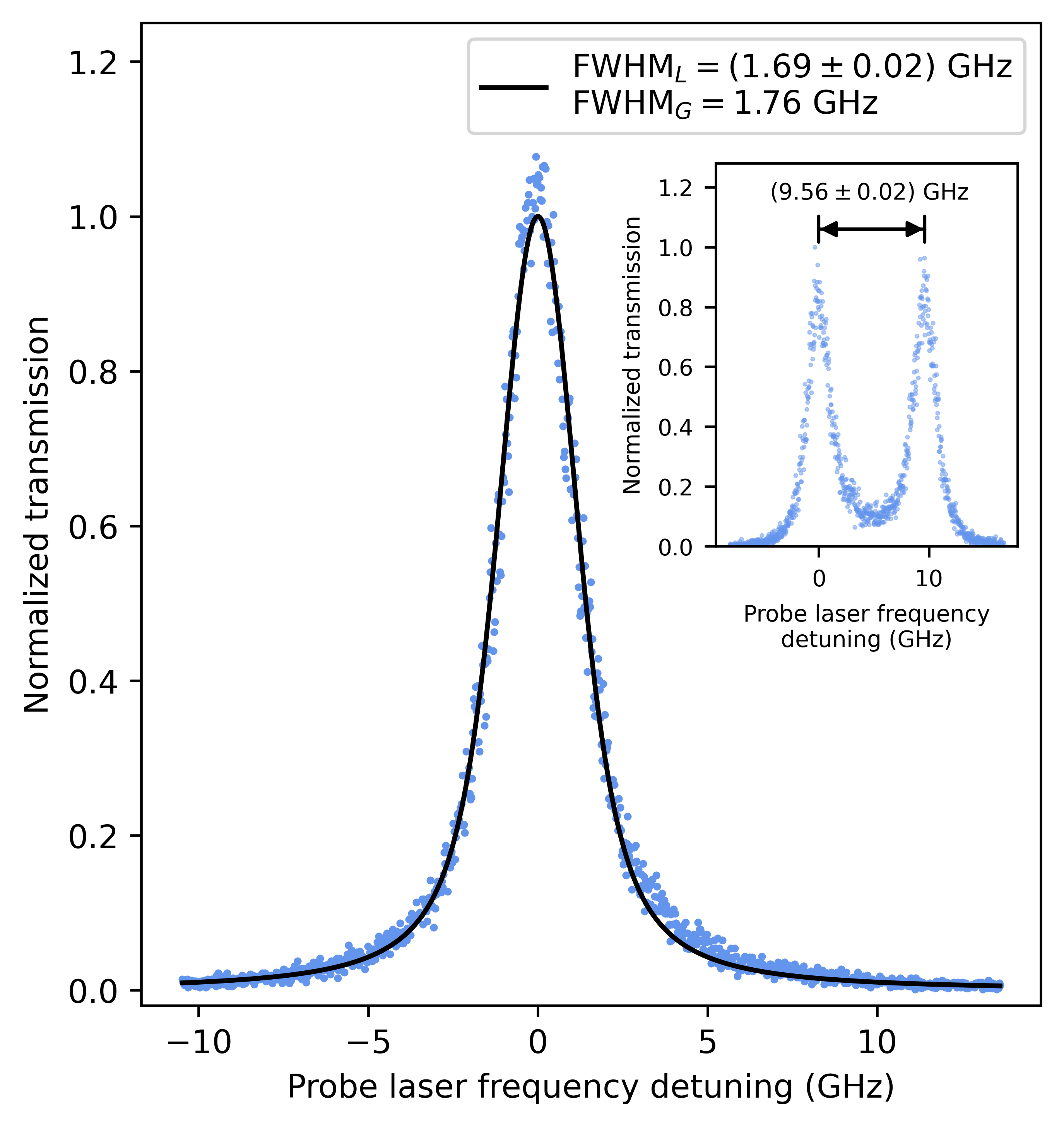}
    \caption{Resonant laser frequency scan over the cavity resonance to determine the cavity linewidth. The polarization of the laser is aligned with the LF cavity mode. The cavity transmission is measured on the ZPL single-photon detector. A Voigt function with a fixed Gaussian $\rm FWHM_G$ contribution, set by the vibration level, is used to fit the data (see text). The fit yields the cavity linewidth $\rm FWHM_L$ as the Lorentzian contribution. The inset shows a laser frequency scan, where a laser polarization is used that illuminates both cavity modes. This measurement reveals a frequency splitting between the LF and HF cavity mode of $\unit[(9.56 \pm 0.02)]{GHz}$.}
    \label{fig:cavity_linewidth}
\end{figure}
\end{center}

To determine the FWHM Lorentzian linewidth of the cavity mode, we scan a laser over the cavity resonance as shown in Fig.~\ref{fig:cavity_linewidth}. In this measurement, the linewidth of the cavity is broadened due to cavity length fluctuations. An independent measurement (as described in Ref.~\cite{herrmann_low-temperature_2024}) yields a vibration level of $\unit[22]{pm}$ root mean square of the cavity length fluctuations. Modeling these vibrations with a Gaussian distribution results in a Gaussian contribution of $\rm FWHM_G=\unit[22]{pm} \times \unit[34]{MHz/pm} \times 2\sqrt{2\ln2}$ next to the Lorentzian cavity linewidth. By fitting a Voigt function with the determined Gaussian contribution, an intrinsic cavity linewidth of $\rm FWHM_L=\unit[(1.69 \pm 0.02)]{GHz}$ is obtained. This corresponds to a cavity quality factor of $280 \times 10^3$ and, for our cavity geometry, to a cavity finesse of $2800$. The cavity mirrors are both coated with dielectric Bragg mirrors (Laseroptik), which are optimized for maximum reflectivity at $\unit[637]{nm}$. At this wavelength, the fiber mirror (sample mirror) exhibits a transmission value of $\unit[50]{ppm}$ ($\unit[875]{ppm}$) for air (diamond) termination. With the determined finesse, total cavity losses of $\unit[2260]{ppm}$ are calculated, which reveal additional losses of $\unit[1260]{ppm}$. We attribute these additional losses to residual scattering at the interface of different refractive indices in the microcavity. Considering these losses, an outcoupling efficiency through the sample mirror of about $\unit[39]{\%}$ is estimated, which reflects our asymmetric cavity design.

The inset of Fig.~\ref{fig:cavity_linewidth} shows another resonant laser scan, which is used to determine the frequency splitting between the LF and the HF cavity mode of $\unit[(9.56 \pm 0.02)]{GHz}$. For the used spherical fiber mirror, we do not observe a significant mode splitting on the bare mirror and hence attribute it mainly to the birefringence of the diamond membrane.

\section{NV center PLE measurements\label{apx:PLEs}}
The power values of the lasers that are deployed via the cavity fiber mirror are stated as measured in transmission after the cavity (on cavity resonance), whereas the power values of the lasers deployed via the sample mirror side are measured free-space before entering the cryostat. All PLE measurements of the NV center are performed by repetitively running a certain PLE sequence for every swept excitation laser frequency. The number of repetitions and the exact PLE sequences for the different PLE measurements are outlined in the following subsections. All PLE sequences start with a $\unit[515]{nm}$ repump pulse, which prepares the NV center with high probability in its negatively charged state and predominantly in its $m_s=0$ spin ground state. This pulse consists of a $\unit[50]{\upmu s}$ long and $\unit[60]{\upmu W}$ strong $\unit[515]{nm}$ repump laser pulse followed by a $\unit[5]{\upmu s}$ wait time.

\subsection{PLE sequence of the NV center $E_y$ transition}
In the PLE measurement of the NV center $E_y$ transition, shown in Fig.~\ref{fig:NV_optical_characterization}(a), the following PLE sequence is applied $200 \times 10^3$ times per excitation laser frequency. The sequence starts with the above-described repump pulse followed by a $\unit[100]{\upmu s}$ long $\unit[0.2]{nW}$ spin initialization laser pulse on the $E_1$ transition to initialize the NV center in the $m_s=0$ spin ground state. Then a free-space $\unit[1]{nW}$ excitation laser pulse is applied for $\unit[200]{\upmu s}$, while simultaneously measuring the ZPL signal of the NV center in the cross-polarized detection. The ZPL counts are acquired time-resolved, which results in an exponentially decaying signal as can be seen exemplarily in Fig.~\ref{fig:cw_readout}(a) after averaging the repetitions for each excitation laser frequency. For the used excitation laser power of $\unit[1]{nW}$, the decay is described well by a monoexponential fit on top of a constant offset level, which is determined by excitation laser leakage into the cross-polarized detection. Thus, the monoexponential fit amplitudes resemble a background-corrected measure of the ZPL counts, which are plotted in Fig.~\ref{fig:NV_optical_characterization}(a). In this measurement, the NV center is excited via the HF cavity mode, and fluorescence is detected in the LF cavity mode. In addition, the LF mode is kept on resonance with the excitation laser frequency by the cavity lock.

\subsection{PLE sequence of the NV center $E_1$ and $E_2$ transition}
\begin{center}
\begin{figure}[ht]
    \includegraphics[width=\linewidth]{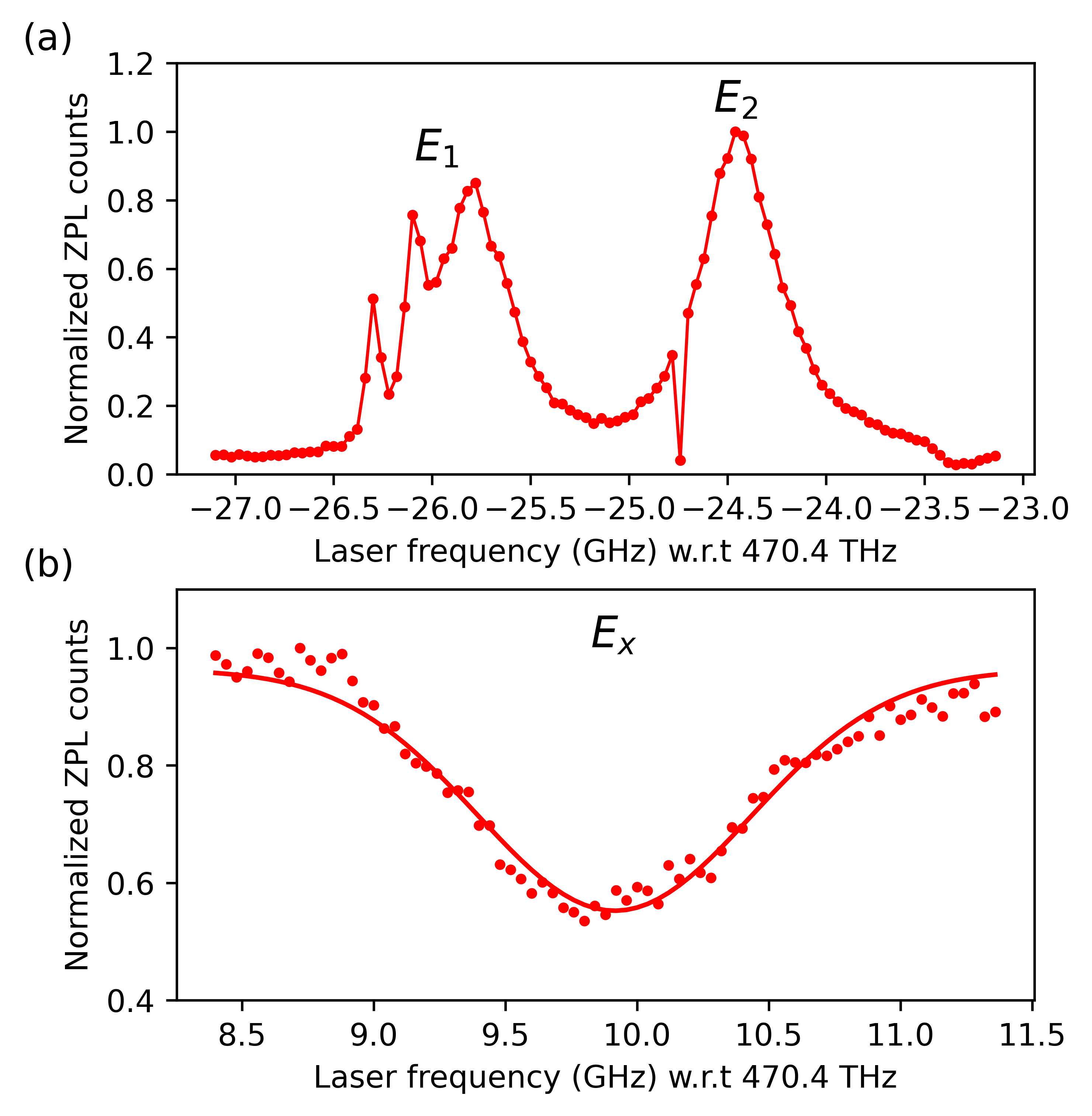}
    \caption{Further PLE measurements of the NV center used in the main text. (a) NV center $E_1$ and $E_2$ transition. For the laser frequency of $\unit[-24.74]{GHz}$, we suspect that our spin initialization laser did not switch on, resulting in ZPL counts at the background level. (b) NV center $E_x$ transition with a Gaussian fit as a guide for the eye.}
    \label{fig:ples}
\end{figure}
\end{center}

In the PLE measurement of the NV center $E_1$ and $E_2$ transition, shown in Fig.~\ref{fig:ples}(a), the following PLE sequence is applied $100 \times 10^3$ times per excitation laser frequency. The sequence starts with the above-described repump pulse followed by $300$ short excitation pulses on the NV center $E_y$ transition to prepare the NV center in the $m_s=\pm1$ spin ground states (see Fig.~\ref{fig:spin_pump}(a)). Then a $\unit[0.01]{nW}$ excitation laser pulse is applied for $\unit[100]{\upmu s}$ via the cavity fiber followed by $30$ short excitation pulses to read out the NV center $m_s=0$ spin state (as used in the main text). If the excitation laser is resonant with a transition associated with the $m_s=\pm1$ spin states, some population is optically pumped back into the $m_s=0$ spin ground state, and the ZPL signal is detected by the readout as shown in Fig.~\ref{fig:ples}(a). This technique is used since the direct ZPL signal acquisition of the weaker coupled NV transitions in cross-polarized resonant excitation and detection is inefficient. In this measurement, the LF cavity mode is used for readout and kept on resonance with the NV center $E_y$ transition by the cavity lock.

In the acquired PLE measurement shown in Fig.~\ref{fig:ples}(a), two additional peaks with a splitting of about $\unit[200]{MHz}$ that overlap with the $E_1$ transition are observed. We attribute these peaks to the transitions between the Zeeman-split $m_s=\pm1$ ground states and the excited $E_y$ state. This observation reveals a finite strength of these transitions for the investigated NV center.

\subsection{PLE sequence of the NV center $E_x$ transition}
In the PLE measurement of the NV center $E_x$ transition, shown in Fig.~\ref{fig:ples}(b), the following PLE sequence is applied $100 \times 10^3$ times per excitation laser frequency. The sequence starts with the above-described repump pulse. Then a $\unit[200]{\upmu s}$ long $\unit[50]{nW}$ excitation laser pulse is applied via the cavity fiber, followed by $30$ short excitation pulses to read out the NV center $m_s=0$ spin state. If the excitation laser is resonant with a transition associated with the $m_s=0$ spin state, some population is optically pumped into the $m_s=\pm1$ spin ground states, and a reduction of ZPL signal is detected by the readout as shown in Fig.~\ref{fig:ples}(b). This technique is used since the direct ZPL acquisition of the weaker coupled NV transitions in cross-polarized resonant excitation and detection is inefficient. In this measurement, the LF cavity mode is used for readout and kept on resonance with the NV center $E_y$ transition by the cavity lock.

\section{Pulsed resonant excitation\label{apx:excitation_pulse}}
\begin{center}
\begin{figure}[ht]
    \includegraphics[width=\linewidth]{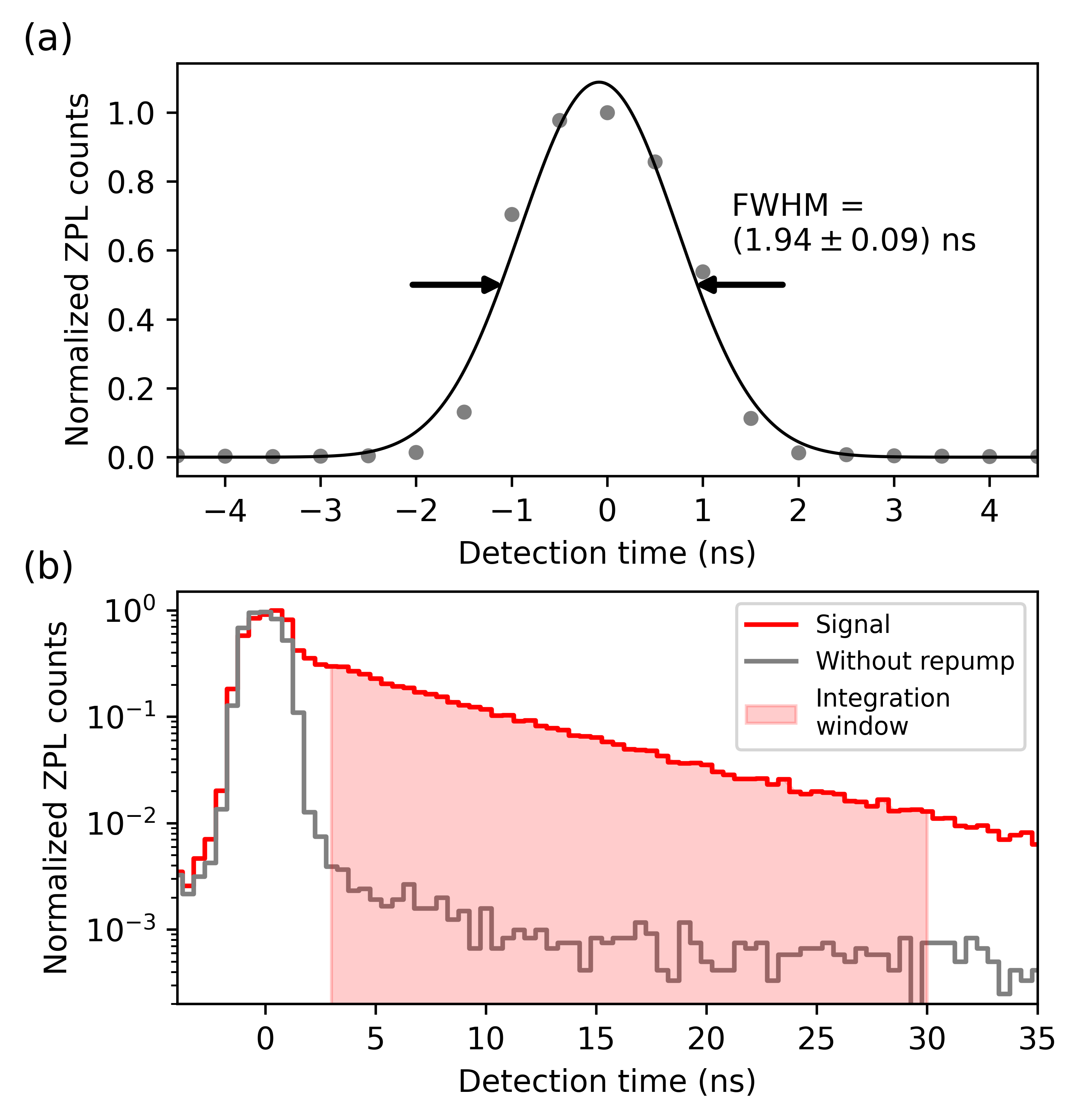}
    \caption{(a) Temporal shape of the short excitation pulse used in this study. (b) Short excitation pulse next to a time-resolved NV center readout signal. For the short excitation pulse measurement, the NV center is 'switched off' by not applying the charge repump pulse at the start of the measurement sequence.}
    \label{fig:pulse}
\end{figure}
\end{center}

Figure~\ref{fig:pulse}(a) shows the temporal shape of the short excitation pulse used throughout the study. A FWHM value of $\unit[(1.94\pm0.09)]{ns}$ is determined by a Gaussian fit. In addition, Fig.~\ref{fig:pulse}(b) shows the same short excitation pulse next to a time-resolved pulsed readout measurement of the NV center $E_y$ transition. The short excitation pulse is declined before the integration window, from $\unit[3]{ns}$ to $\unit[30]{ns}$ with respect to the excitation pulse center, starts. The integration window includes about $\unit[75]{\%}$ of the studied Purcell-enhanced NV center emission, assuming the NV center population inversion is completed $\unit[1]{ns}$ after the excitation pulse center.

\section{NV center excited state lifetime measurements\label{apx:lifetime}}
\begin{center}
\begin{figure}[ht]
    \includegraphics[width=\linewidth]{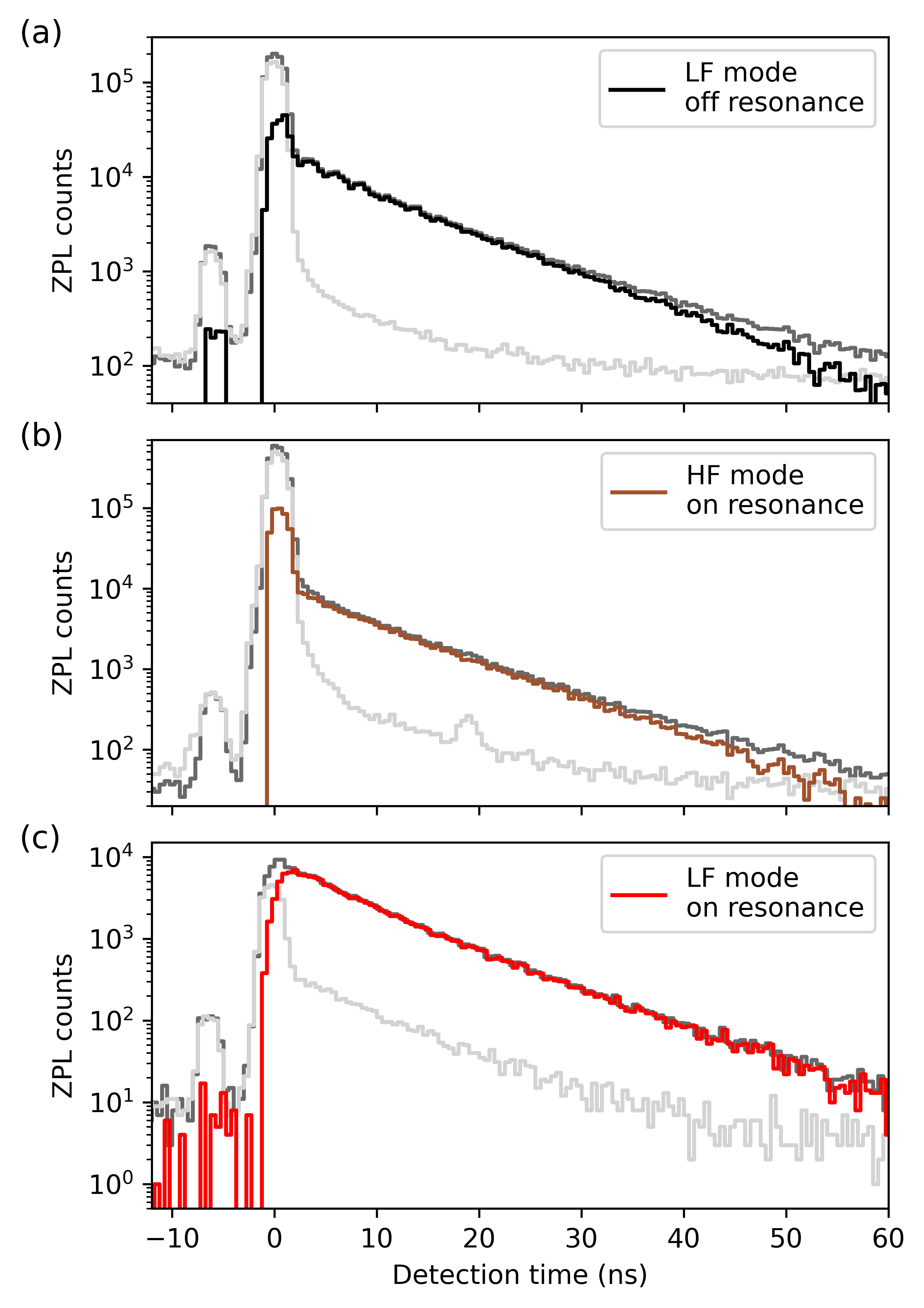}
    \caption{Background correction of the NV center $E_y$ excited state lifetime measurement. Data in dark (light) gray shows the ZPL signal of the first (last) $5$ short excitation pulses of the total applied $100$ pulses. The small peak at a detection time of about $\unit[-5]{ns}$ is light of the excitation pulse that is backscattered into the free-space detection before reaching the cryostat. (a) Background correction of the lifetime measurement in the LF mode with a cavity detuning of $\unit[-4]{GHz}$ from the $E_y$ transition. (b) Background correction of the lifetime measurement in the HF mode and the cavity on resonance with the $E_y$ transition. (c) Background correction of the lifetime measurement in the LF mode and the cavity on resonance with the $E_y$ transition.}
    \label{fig:lifetime_bg_correction}
\end{figure}
\end{center}

The NV center excited state lifetime measurement sequences used in Fig.~\ref{fig:NV_optical_characterization}(c) start with a $\unit[515]{nm}$ repump pulse, which prepares the NV center with high probability in its negatively charged state. This pulse consist of a $\unit[50]{\upmu s}$ long and $\unit[60]{\upmu W}$ strong $\unit[515]{nm}$ repump laser pulse followed by a $\unit[5]{\upmu s}$ wait time. Subsequently, a $\unit[0.2]{nW}$ spin initialization laser pulse is applied for $\unit[100]{\upmu s}$ ($\unit[20]{\upmu s}$) on the $E_1$ transition to initialize the NV center in the LF (HF) cavity mode in its $m_s=0$ spin ground state. Then $100$ short excitation pulses are used in the sequence to excite the NV center, followed by time-resolved detection of the ZPL signal. To correct for background, we make use of the pumping into the NV center $m_s=\pm1$ spin ground states with more applied short excitation pulses (see Fig.~\ref{fig:spin_pump}(a)). This allows us to take the high signal-to-noise part of the first $5$ short excitation pulses and subtract the last $5$, which corrects for constant background contributions. The data underlying the background subtraction is shown in Fig.~\ref{fig:lifetime_bg_correction} and leads to the lifetime measurement data used in the main text Fig.~\ref{fig:NV_optical_characterization}(c).

\begin{center}
\begin{figure}[ht]
    \includegraphics[width=\linewidth]{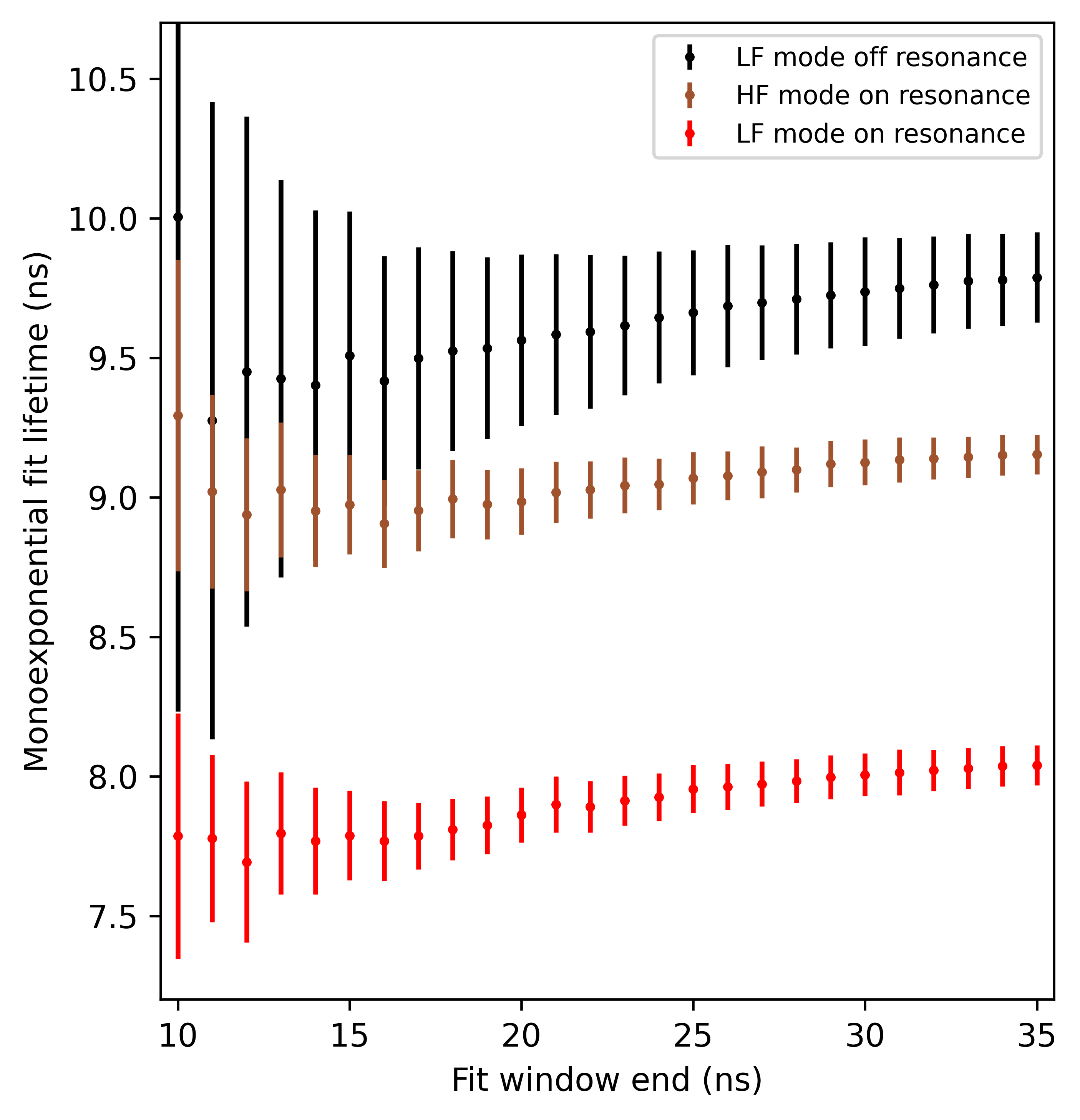}
    \caption{Extracted excited state lifetime from a monoexponential fit on the lifetime measurement data of main text Fig.~\ref{fig:NV_optical_characterization}(c). The fit window start is set to $\unit[5]{ns}$ while the fit window end is varied.}
    \label{fig:lifetime_fit_window}
\end{figure}
\end{center}

Despite the background subtraction, we see that the data deviates from a monoexponential behavior with increasing detection time, which is also observed in the extracted lifetimes for different fit windows shown in Fig.~\ref{fig:lifetime_fit_window}. This hints towards a time-dependent background contribution that might stem from other weaker coupled NV centers in the diamond membrane or fluorescence of the cavity mirrors.

\begin{center}
\begin{figure}[ht]
    \includegraphics[width=\linewidth]{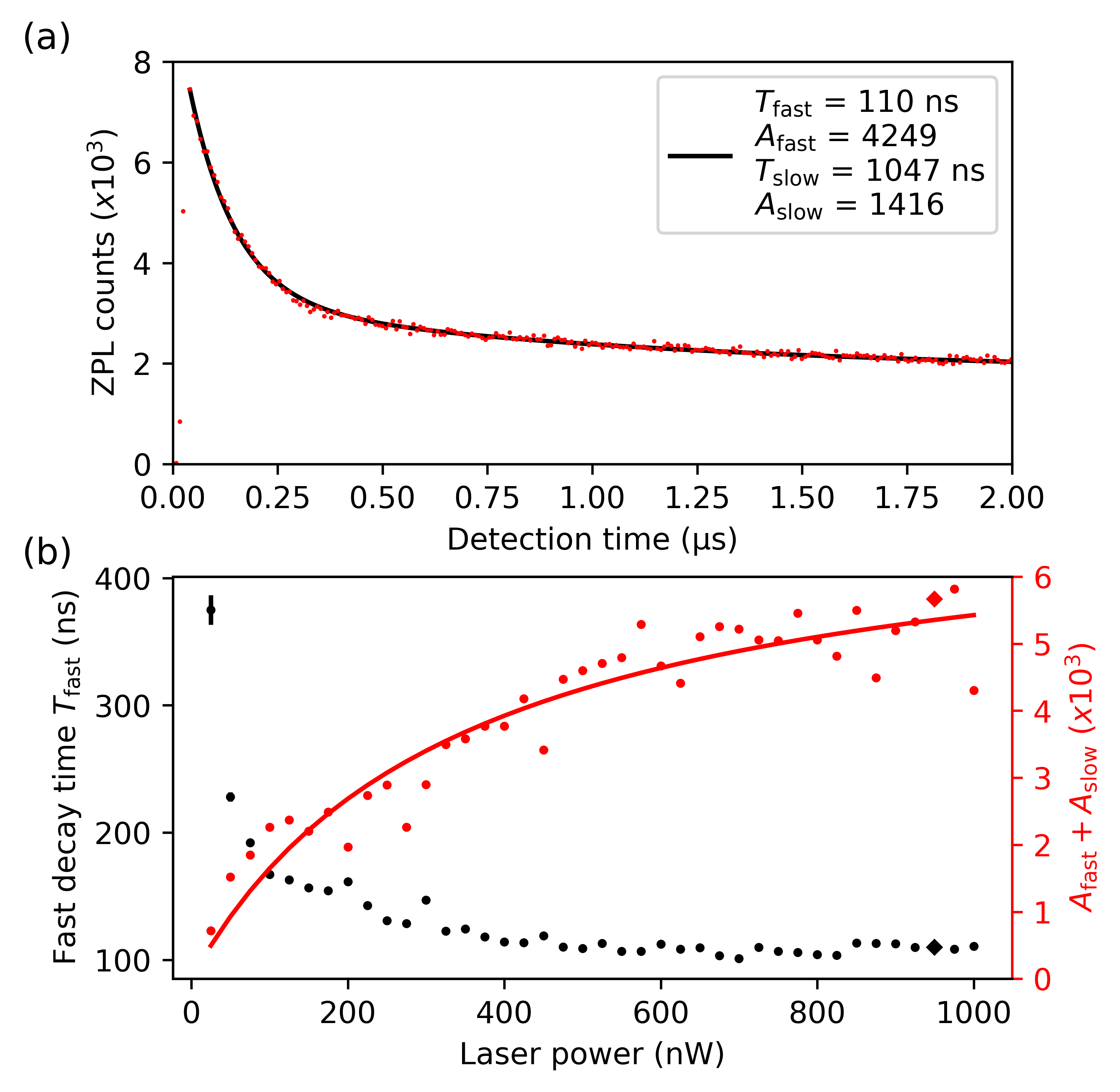}
    \caption{(a) Exemplary time-resolved ZPL signal of the NV center $E_y$ transition saturation measurement. In this measurement, a resonant laser power of $\unit[950]{nW}$ is used, and the black solid line is a double exponential fit with offset to the data. The fast decay for short detection times is described with the time constant $T_\text{fast}$ while the slow decay is captured by $T_\text{slow}$. (b) NV center $E_y$ transition saturation measurement with extracted fit parameters. The total fit amplitude $A_\text{fast}+A_\text{slow}$ resembles the saturating ZPL counts, and $T_\text{fast}$ is the fast exponential decay time constant. The total fit amplitudes that depend on the laser power $P$ are fitted to a function proportional to $P/(P+2P_\text{sat})$ as a guide for the eye. The individual measurement shown in (a) is indicated by the diamond-shaped data point.}
    \label{fig:cw_readout}
\end{figure}
\end{center}

Furthermore, an off resonance lifetime of $\tau_\text{off}=\unit[(9.5\pm0.4)]{ns}$ is measured for the NV center $E_y$ excited state in the main text, which deviates from the expected about $\unit[12.4]{ns}$ lifetime in bulk samples \cite{hermans_entangling_2023}. This reveals a larger excited state decay rate of the studied NV center, which we attribute to strain-induced mixing in the excited states \cite{manson_nitrogen-vacancy_2006,goldman_state-selective_2015}. In Fig.~\ref{fig:cw_readout}(b), we conduct a continuous wave saturation measurement of the NV center $E_y$ transition using $\unit[32]{\upmu s}$ long resonant pulses and detect the ZPL signal time-resolved as shown in Fig.~\ref{fig:cw_readout}(a). The ZPL signal fits well to a double exponential fit with a constant offset accounting for excitation laser leakage into the cross-polarized detection. The summed amplitudes resemble a measure for the saturating ZPL counts, while the exponential time constants give insights into the decay dynamics of the studied NV center. We attribute the fast decay time constant to additional decay from the excited state, next to the decay into the $m_s=0$ spin ground state. Assuming that saturating the NV center with high excitation power leads to an average excited state population of $\unit[50]{\%}$, half of the fitted fast decay time constant $\unit[110]{ns}/2$ (see Fig.~\ref{fig:cw_readout}(b)) determines the additional decay rate. With that additional decay, the predicted lifetime of the excited state is about $\unit[10]{ns}$, which matches with the value of the off resonance lifetime measurement.

\section{NV center spin initialization and readout\label{apx:init_and_readout}}
All measurement sequences used in Fig.~\ref{fig:NV_spin_characterization} and Fig.~\ref{fig:spin_photon_correlations} start with a $\unit[515]{nm}$ repump pulse, which prepares the NV center with high probability in its negatively charged state. This pulse consist of a $\unit[50]{\upmu s}$ long and $\unit[60]{\upmu W}$ strong $\unit[515]{nm}$ repump laser pulse followed by a $\unit[5]{\upmu s}$ wait time. Subsequently, and if not stated otherwise, a $\unit[100]{\upmu s}$ long $\unit[0.2]{nW}$ spin initialization laser pulse on the $E_1$ transition is applied to initialize the NV center in its $m_s=0$ spin ground state. Then the actual measurement sequence, composed of microwave (and individual short resonant excitation) pulses, is played, followed by a series of short excitation pulses to read out the NV center qubit state. If at least one photon is detected during the readout, the $m_s=0$ qubit state is measured; otherwise, the $m_s=-1$ qubit state is assigned. Since these outcomes are subject to measurement errors like photon loss as well as dark or background photon detection events, a readout correction is used to recover the measurement statistics. For this purpose, the initialization fidelity in the $m_s=0$, as well as $m_s=\pm1$ spin state and their corresponding readout fidelities, are determined, and the correction formalism as outlined in Ref.~\cite{pompili_multi-node_2021} is used.

\begin{center}
\begin{figure}[ht]
    \includegraphics[width=\linewidth]{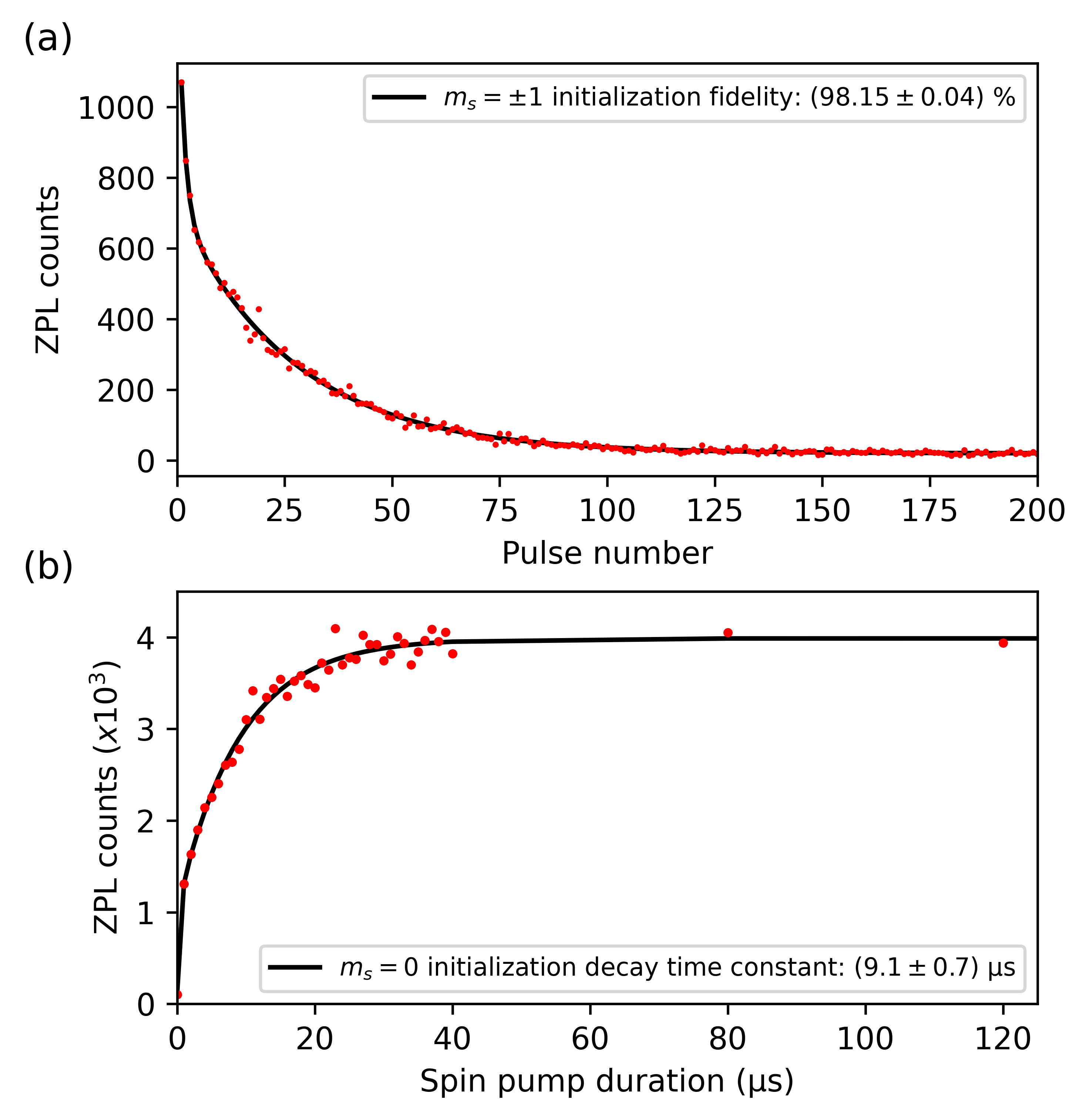}
    \caption{(a) Spin initialization measurement in the $m_s=\pm1$ spin states by optical pumping with short excitation pulses on the NV center $E_y$ transition. Before the short excitation pulses are applied, a $\unit[0.2]{nW}$ spin initialization pulse is applied for $\unit[120]{\upmu s}$ on the $E_1$ transition. (b) Indirect spin initialization measurement in the $m_s=0$ spin state by optical pumping on the NV center $E_1$ transition and readout by $5$ short excitation pulses. In this measurement, a spin initialization pulse power of $\unit[0.2]{nW}$ is used, while the spin initialization pulse duration is varied.}
    \label{fig:spin_pump}
\end{figure}
\end{center}

The initialization in the $m_s=\pm1$ spin states is performed using the short excitation pulses that are resonant with the NV center readout transition. The detected ZPL counts in the integration window of $\unit[3]{ns}$ to $\unit[30]{ns}$ after each short excitation pulse are shown in Fig.~\ref{fig:spin_pump}(a). The ZPL counts, depending on the pulse number, follow a double exponential behavior similar to the second-order correlation measurement in Fig.~\ref{fig:NV_optical_characterization}(d). The measurement shows that after about $200$ short excitation pulses, a constant ZPL count level is reached. Comparing the total ZPL count amplitude with the offset determined by fitting an initialization fidelity of $\unit[(98.15\pm0.04)]{\%}$ is determined.

The initialization in the $m_s=0$ spin state is performed by optical pumping on the NV center $E_1$ transition. To investigate the spin initialization behavior, the NV center is first prepared in the $m_s=\pm1$ spin states by $600$ short excitation pulses, and then a $\unit[0.2]{nW}$ spin initialization laser pulse is applied for different durations, followed by $5$ short excitation pulses for readout. The measured readout ZPL counts depending on the spin initialization pulse duration are shown in Fig.~\ref{fig:spin_pump}(b). For longer spin initialization pulse durations, the detected ZPL counts increase until they settle exponentially. From a double exponential fit, a slow decay time constant of $\unit[(9.1\pm0.7)]{\upmu s}$ is determined. This indirect spin initialization measurement via the NV center readout transition gives insights about the required spin initialization pulse durations for the used laser power, but does not allow us to determine an initialization fidelity in the $m_s=0$ spin state.

To estimate the $m_s=0$ initialization fidelity, the NV center $m_s=0$ spin state is first heralded by measuring a photon after a single short excitation pulse and then read out with $200$ short excitation pulses. This results in the maximal readout probability, since the $m_s=0$ spin state is heralded just before the readout. After this first readout, further short excitation pulses are used to initialize the NV center in the $m_s=\pm1$ states followed by a $\unit[0.2]{nW}$ spin initialization laser pulse on the $E_1$ transition for $\unit[100]{\upmu s}$ and another readout with $200$ short excitation pulses. This results in a second readout probability, which is set by the optical pumping performance of the initialization laser pulse. Comparing the second to the first readout yields an initialization fidelity in the $m_s=0$ spin state of $\unit[(93.5\pm 0.9)]{\%}$. To exclude NV center ionization, a third iteration is performed, leading to the same result as the second readout.

Moreover, this experiment reveals the combined probability that the initial repump and spin initialization pulse prepares the NV center in its negatively charged and $m_s=0$ state by comparing the probability of the first readout with and without the described heralding. From this, a probability of $\unit[(72.8\pm0.6)]{\%}$ is found, which is considered in the readout of the heralded measurements in Fig.~\ref{fig:spin_photon_correlations}. Further, we note that an initialization laser pulse duration of $\unit[20]{\upmu s}$ instead of $\unit[100]{\upmu s}$ is used in the GHZ state measurement of Fig.~\ref{fig:spin_photon_correlations}(c).

\section{NV center Ramsey measurement\label{apx:ramsey_analysis}}
The Ramsey fringes measured in Fig.~\ref{fig:NV_spin_characterization}(c) are determined by the hyperfine coupling of the nitrogen-14 nuclear spin $\omega_\text{hf}$, the free-induction decay time $T_2^*$, and the used artificial detuning $\Delta$. The artificial detuning is implemented by applying a phase of $\tau\Delta$ to the second $\pi/2$ pulse in the Ramsey sequence. The data is well fitted by the following function 
\begin{align}
    P_{m_s=0}(\tau) = c - &\exp\left( -\left(\frac{\tau}{T_2^*}\right)^n\right) \nonumber\\
    &\cdot \sum_{k=-1,0,+1} P_k \cos\left((\Delta + k \omega_{\text{hf}})\tau + \phi\right),
\end{align}
with the offset $c$, the decay exponent $n$, the nitrogen nuclear spin amplitudes $P_k$ and a common phase $\phi$. The amplitudes $P_k$ take partial polarization of the nitrogen nuclear spin into account, and the common phase $\phi$ accounts for a phase offset between the two applied $\pi/2$ pulses.

\section{Bell state spin-photon correlations with spin qubit readout in X-basis \label{apx:spin_photon_correlations_x_basis}}
\begin{center}
\begin{figure}[ht]
    \includegraphics[width=\linewidth]{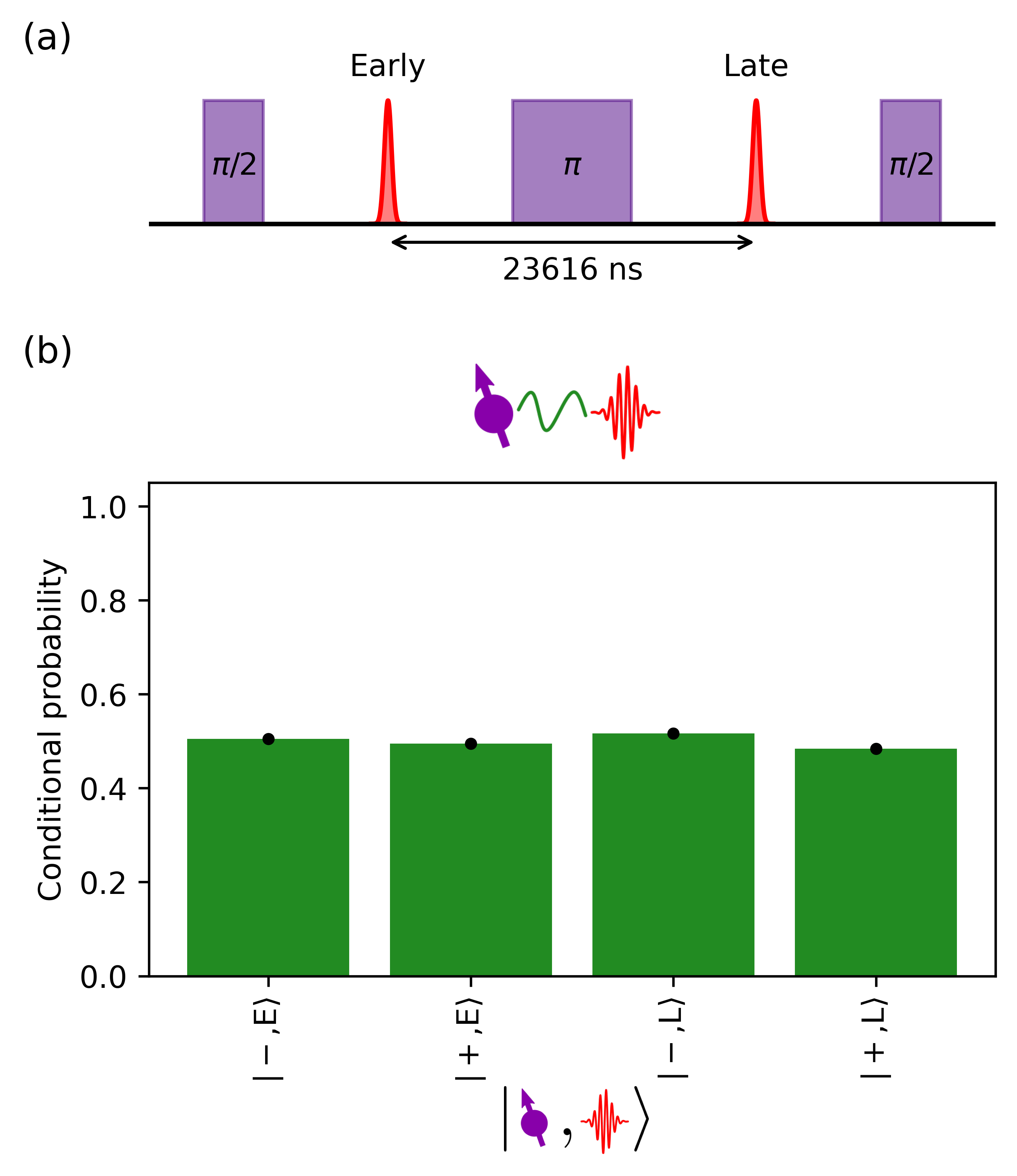}
    \caption{(a) Pulse sequence to generate a spin-photon Bell state and a spin qubit readout in the X-basis. In the experiment, a microwave decoupling interpulse delay time of $\tau/2 = \unit[23568]{ns}$ is used. (b) Conditional probabilities of the spin qubit readout in the X-basis after successful heralding an early ($\ket{\text{E}}$) or late ($\ket{\text{L}}$) photon of the spin-photon Bell state. The error bars are within the black dot size.}
    \label{fig:spin_photon_correlations_x_basis}
\end{figure}
\end{center}

In addition to the Bell state spin-photon correlation measurement of Fig.~\ref{fig:spin_photon_correlations}(b), where the spin qubit is read out in its Z-basis, we perform an X-basis spin qubit readout in Fig.~\ref{fig:spin_photon_correlations_x_basis}(b). The used pulse sequence is depicted in Fig.~\ref{fig:spin_photon_correlations_x_basis}(a), which exhibits another $\pi/2$ pulse after the Bell state generation to perform the qubit readout in the X-basis. As in the main text, $200$ short excitation pulses are used for readout.

In the X-basis measurement, the spin is measured with equal probability in one of the qubit states. In total we record $24456$ photon heralding events in $\unit[5\times10^6]{}$ attempts, which corresponds to a probability of $\unit[0.49]{\%}$ per attempt.

\section{Additional data of a second cavity-coupled NV center\label{apx:second_nv}}
\begin{center}
\begin{figure}[ht]
    \includegraphics[width=\linewidth]{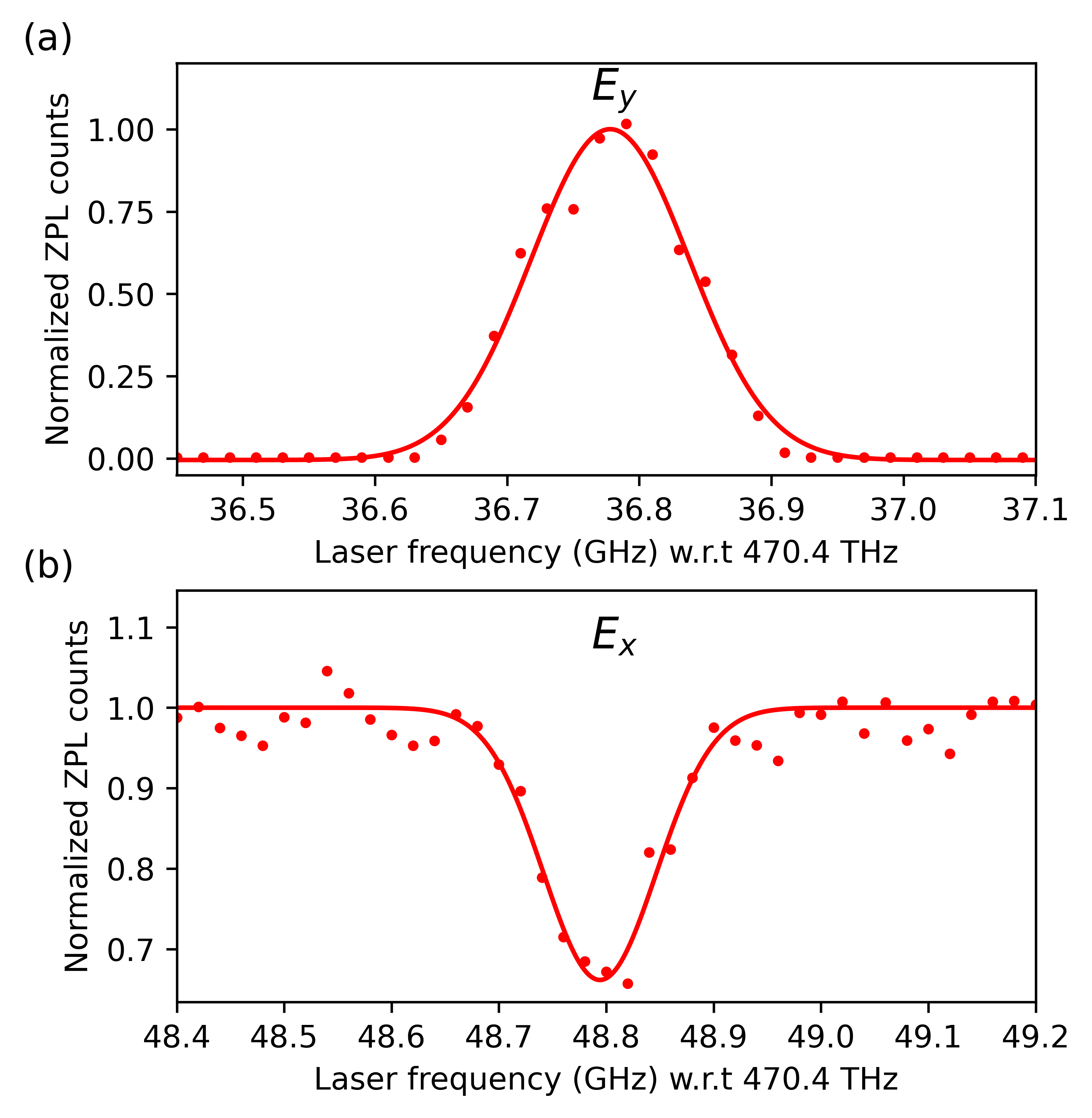}
    \caption{PLE measurements of a second cavity-coupled NV center at a different lateral position on the diamond membrane. (a) NV center $E_y$ transition. (b) NV center $E_x$ transition. The PLE measurements are acquired with similar sequences as for the NV center studied in the main text. The readout is performed with $200$ short excitation pulses.}
    \label{fig:second_nv_ples}
\end{figure}
\end{center}

At a different lateral position on the diamond membrane, a second NV center coupled to the cavity is investigated. For this cavity position a frequency splitting of the two polarization cavity modes of about $\unit[10.7]{GHz}$ is observed and a NV center coupled with its $E_y$ transition at a frequency of about $\unit[36.8]{GHz}$ (with respect to $\unit[470.4]{THz}$) is found. The $E_x$ transition of the NV center is identified with a frequency of about $\unit[48.8]{GHz}$ resulting in a lower lateral strain with $E_x~\text{-}~E_y$ excited state splitting of $\sim \unit[12]{GHz}$ compared to the NV center studied in the main text. The corresponding PLE measurements are shown in Fig.~\ref{fig:second_nv_ples}.

\begin{center}
\begin{figure}[ht]
    \includegraphics[width=\linewidth]{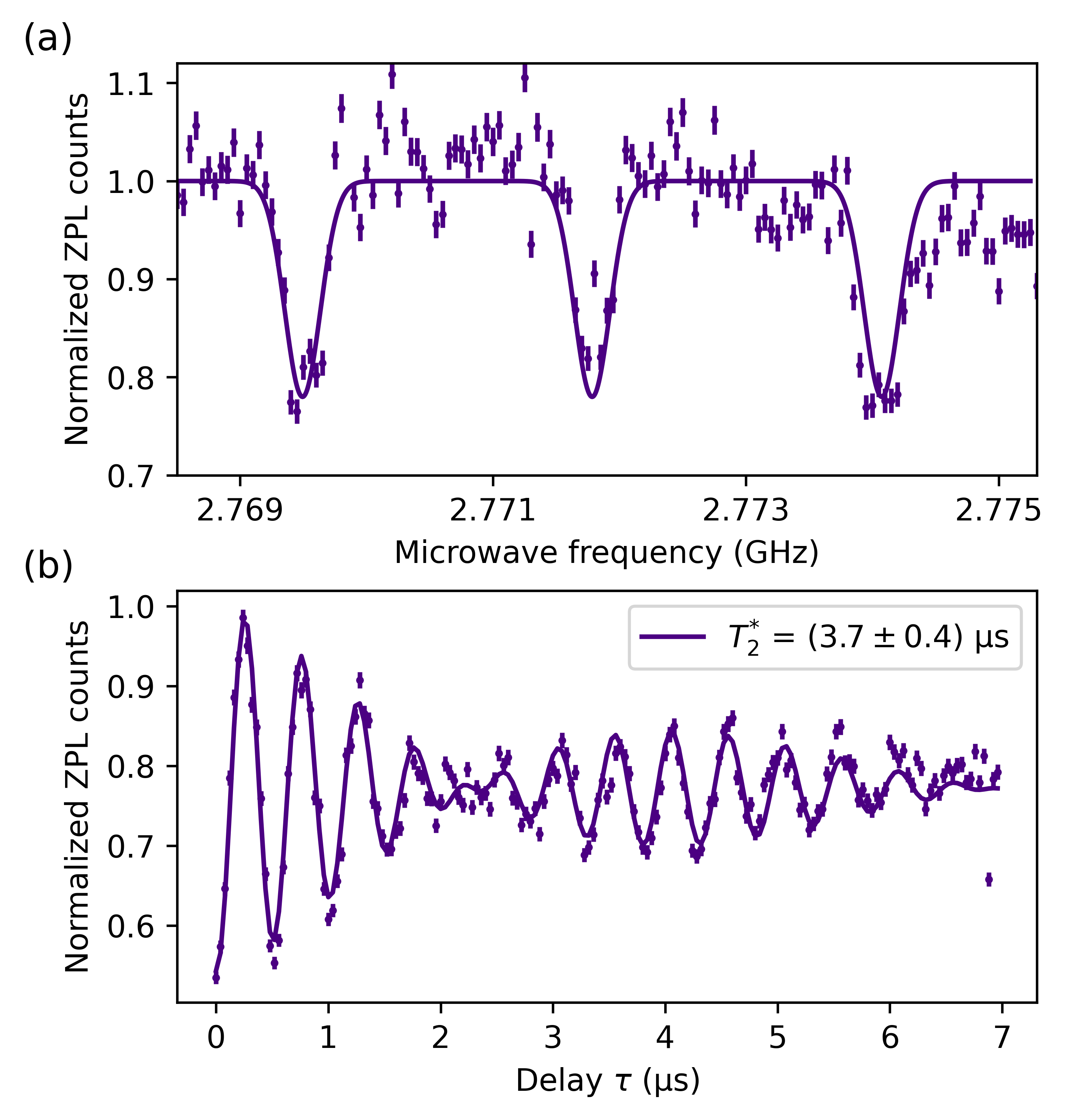}
    \caption{(a) Electron spin resonance measurement. The NV center is initialized in its $m_s=0$ state, then a microwave pulse of $\unit[10]{\upmu s}$ is applied, and finally the NV center $m_s=0$ state is read out with $50$ short excitation pulses. The microwave frequency is swept over the $m_s=-1$ electron spin state, and for each nitrogen nuclear spin resonance ($m_I=-1,0,+1$), a decrease in ZPL counts is observed. The fit with three Gaussian dips is a guide for the eye. (b) Ramsey experiment at the $m_I = 0$ nitrogen nuclear spin resonance frequency with an artificial detuning of $\unit[2]{MHz}$. In this measurement, a microwave $\pi/2$ pulse duration of $\unit[350]{ns}$ is used.}
    \label{fig:second_nv_spin}
\end{figure}
\end{center}

In addition, the spin coherence of this NV center is measured in a Ramsey experiment as shown in Fig.~\ref{fig:second_nv_spin}(b). Compared to the Ramsey experiment shown in Fig.~\ref{fig:NV_spin_characterization}(c) of the main text, only the $m_I = 0$ spin resonance of the coupled nitrogen nuclear spin is driven. This transition is identified beforehand in an electron spin resonance measurement as shown in Fig.~\ref{fig:second_nv_spin}(a). The resulting Ramsey fringes are well fitted by the function
\begin{align}
    P_{m_s=0}(\tau) = c - A~&\exp\left( -\left(\frac{\tau}{T_2^*}\right)^n\right) \nonumber\\
    &\cdot \sum_{k=-1,+1}\cos\left((\Delta + k \omega_{\text{c}}/2)\tau + \phi\right),
\end{align}
with the amplitude $A$ and the carbon-13 nuclear spin coupling frequency $\omega_\text{c}$. Fitting the Ramsey experiment yields a free-induction decay time of $T_2^*=\unit[(3.7\pm0.4)]{\upmu s}$ with a decay exponent of $n=1.1\pm0.3$ and a beating with a frequency of $\unit[(218\pm4)]{kHz}$, which we attribute to the coupling of a carbon-13 nuclear spin. To summarize, NV centers with electron spin coherence times similar to diamond bulk samples can be observed \cite{bernien_heralded_2013}.

\section{Summary of system parameters\label{apx:parameters}}
Next to the parameters of the emitter-cavity system, the efficiency of the free-space cross-polarization setup determines the total efficiency of the resonant excitation and detection scheme involving the short excitation pulses. The photon collection efficiency into the ZPL single-mode fiber comprising all the optics after the microcavity amounts to about $\unit[39]{\%}$, and the used single-photon detectors are specified with $\unit[70]{\%}$. Including the integration time window used for short excitation pulse readout with a collected fraction of about $\unit[75]{\%}$, the total ZPL setup efficiency is estimated to be about $\unit[20]{\%}$. Considering the ZPL emission into the LF mode of about $\unit[18]{\%}$ together with the cavity outcoupling efficiency of $\unit[39]{\%}$, a detector click probability of about $\unit[1.4]{\%}$ per pulse is estimated upon NV center excitation as stated in the main text.

An overview of measured, estimated, and simulated values of this work is summarized in Table \ref{tab:summary}.

\begin{table}[ht]
\caption{Overview of system parameters.\label{tab:summary}}
\footnotesize
\begin{ruledtabular}
\begin{tabular}{ll}
\textrm{Parameter}&
\textrm{Value}\\
\colrule
Fiber mirror radius of curvature & $\unit[21.4]{\upmu m}$ \\
Cavity air gap & $\unit[6.39]{\upmu m}$ \\
Diamond thickness & $\unit[6.20]{\upmu m}$ \\
Hybrid cavity mode number $q$ & $67$ \\
Effective cavity length $L_\text{eff}$ \cite{van_dam_optimal_2018} & $\unit[13.2]{\upmu m}$ \\
Estimated cavity beam waist $\omega_0$ \cite{van_dam_optimal_2018} & $\unit[1.46] {\upmu m}$\\
Estimated cavity mode volume $V$ \cite{van_dam_optimal_2018} & $\unit[86]{\lambda^3}$\\
Cavity Lorentzian linewidth $\kappa/2\pi$ &  $\unit[(1.69\pm0.02)]{GHz}$ \\
Root mean square cavity length fluctuations & $\unit[22]{pm}$ \\
Cavity mode dispersion slope & $\unit[34]{MHz/pm}$ \\
Cavity quality factor $Q$ & $280 \times 10^3$ \\
Cavity finesse $\finesse$  &  $2800$ \\
Cavity outcoupling efficiency& $\unit[39]{\%}$\\
Calculated cavity transmission & $\unit[8]{\%}$\\
Calculated vibration-averaged Purcell factor \cite{herrmann_coherent_2024}& $12$\\
NV center Debye-Waller factor $\beta_0$ & $0.03$ \cite{riedel_deterministic_2017}\\
Measured Purcell factor LF mode $F_{P,\text{LF}}$& $7.3 \pm 1.6$ \\
Measured Purcell factor HF mode $F_{P,\text{HF}}$& $2.0 \pm 1.4$ \\
Measured cooperativity LF mode $\beta_0 F_{P,\text{LF}}$& $0.22 \pm 0.05$\\
Measured cooperativity HF mode $\beta_0 F_{P,\text{HF}}$& $0.06 \pm 0.04$\\
ZPL emission into LF mode $\beta_\text{LF}$& $\unit[18]{\%}$\\
Cavity outcoupled ZPL emission LF mode& $\unit[7]{\%}$\\
ZPL detector click probability per pulse LF mode & $\unit[1.4]{\%}$\\
(end-to-end efficiency at saturation)
\end{tabular}
\end{ruledtabular}
\end{table}

\clearpage
\bibliography{bib}

\end{document}